\documentclass[bibyear]{aa}

\usepackage{graphicx}
\usepackage{natbib}
\usepackage{float}
\usepackage{color}
\usepackage{txfonts}
\usepackage{ulem}
\begin{document} 

   \title{LOFAR observations of PSR B0943+10: profile evolution and discovery of a systematically changing profile delay in Bright mode}

   \author{A.~V.~Bilous
          \inst{1}
          \and 
          J.~W.~T.~Hessels
          \inst{2,3}
          \and
          V.~I.~Kondratiev
          \inst{2,6}
          \and
          J.~van~Leeuwen
          \inst{2,3}
          \and
          B.~W.~Stappers
          \inst{4}
          \and
          P.~Weltevrede
          \inst{4}
          \and
          H.~Falcke
          \inst{1,2}
          \and
          T.~E.~Hassall
          \inst{5}
          \and
          M.~Pilia
          \inst{2}
          \and
          E.~Keane
          \inst{7,8}
          \and
          M.~Kramer
          \inst{4,9}
          \and
          J.-M.~Grie\ss{}meier
          \inst{10,11}
          \and
          M.~Serylak
          \inst{12}
          }


    \institute{Department of Astrophysics/IMAPP,
              Radboud University Nijmegen,
              P.O. Box 9010,
              6500 GL Nijmegen,
              The Netherlands\\
              \email{a.bilous@science.ru.nl}
         \and
             ASTRON, the Netherlands Institute for Radio Astronomy, Postbus 2,
             7990 AA, Dwingeloo, The Netherlands 
         \and 
             Anton Pannekoek Institute for Astronomy, University of Amsterdam,
             Science Park 904, 1098 XH Amsterdam, The Netherlands
         \and
             Jodrell Bank Centre for Astrophysics, School of Physics and Astronomy, 
             University of Manchester, Manchester M13 9PL, UK
         \and
             School of Physics and Astronomy, University of Southampton, 
             Southampton, SO17 1BJ, UK
         \and
             Astro Space Centre of the Lebedev Physical Institute, 
             Profsoyuznaya str. 84/32, Moscow 117997, Russia
         \and
             Centre for Astrophysics and Supercomputing, Swinburne
             University of Technology, Mail H30, PO Box 218, VIC 3122, Australia 
         \and
            ARC Centre of Excellence for All-sky Astrophysics (CAASTRO)
         \and
            MPI f\"{u}r Radioastronomy, Auf dem H\"{u}gel 69, 53121 Bonn, Germany
         \and
            LPC2E - Universit\'{e} d'Orl\'{e}ans / CNRS, France
         \and 
            SR de Nan\c{c}ay, Observatoire de Paris - CNRS/INSU, 
            USR 704 - Univ. Orl\'{e}ans, OSUC, route de Souesmes, 18330 Nan\c{c}ay, France
         \and
            Department of Astrophysics, University of Oxford, Denys Wilkinson Building, Keble Road, Oxford OX1 3RH, UK}

\abstract{
We present broadband, low-frequency (25$-$80\,MHz and 110$-$190\,MHz) LOFAR
observations of  PSR~B0943+10, with the goal of better illuminating the
nature of its enigmatic mode-switching behaviour.  This pulsar shows
two relatively stable states: a Bright (B) and Quiet (Q) mode, each
with different characteristic brightness, profile morphology, and
single-pulse properties.  We model the average profile evolution both in frequency
and time from the onset of each mode, and highlight the differences
between the two modes.  In both modes, the profile evolution can be
well explained by radius-to-frequency mapping at altitudes within a
few hundred kilometres of the stellar surface.  If both B and Q-mode
emission originate at the same magnetic latitude, then we find that the
change of emission height between the modes is less than 6\%.  We
also find that, during B-mode, the average profile is gradually shifting towards later spin
phase and then resets its position at the next Q-to-B transition.  The observed B-mode profile delay
 is frequency-independent (at least from 25$-$80\,MHz) and asymptotically changes towards a stable value
of about $4\times10^{-3}$ in spin phase by the end of mode instance, much too
large to be due to changing spin-down rate. 
Such a delay can be interpreted as a gradual movement of 
the emission cone against the pulsar's
direction of rotation, with different 
field lines being illuminated over time. Another interesting explanation
is a possible variation of accelerating potential inside the polar gap. This explanation
connects the observed profile delay to the gradually evolving subpulse drift rate, which
depends on the gradient of the potential across the field lines. }
   

\titlerunning{LOFAR observations of PSR B0943+10}
\maketitle
%

\section{Introduction}
\label{sec:intro}

PSR~B0943+10, hereafter B0943, is an old ($\tau_c=5\times10^6$ years),
slow ($P=1.098$\,s) pulsar that has two distinct and reproducible modes of
radio emission. The modes show two distinct average pulse profiles, and are named after their phase-integrated
fluxes: in the Bright (B) mode the pulsar is about two times brighter than
in the Quiet (Q) mode (at 62 and 102\,MHz, \citealp{Suleymanova1984}).

\begin{figure*}[ht]
   \centering
 \includegraphics[scale=0.8]{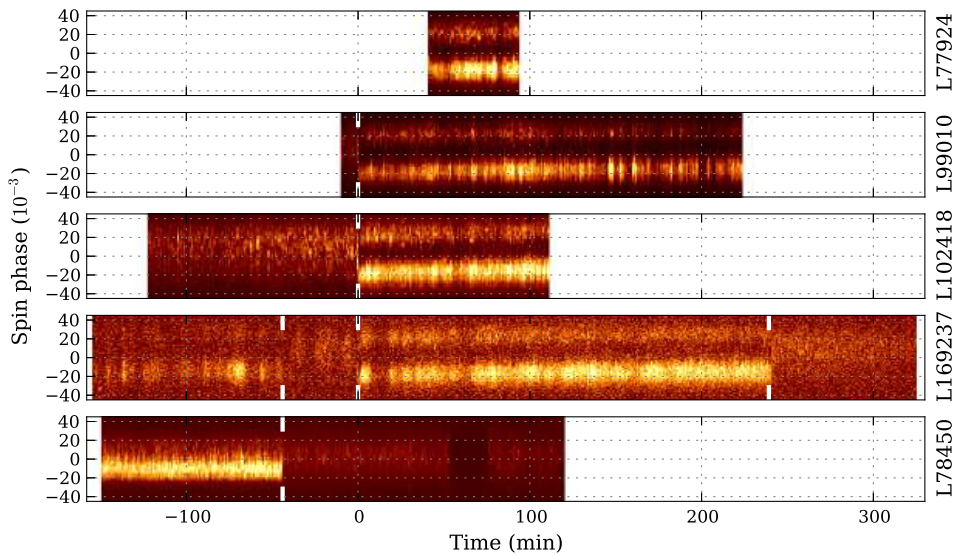}

 \caption{One-minute sub-integrations versus rotational phase for five
   observing sessions.  The top four panels show LBA (25$-$80\,MHz)
   observations; the bottom one used the HBAs
   (110$-$190\,MHz). Observation IDs are included on the right of each
   subplot. Mode transitions are marked with white ticks.  Here and
   throughout the paper the three central LBA observations are aligned
   at the transition from Q-to-B, with the goal to facilitate visual
   comparison of time-dependent emission parameters.  The B-mode of
   the observation L77924 was aligned relative to the others according
   to the ratio of component peak heights (see Fig.~\ref{fig:irat}). The HBA
   observation was aligned by the Q-to-B transition in observation
   L169237. On the Y-axis, the pulsar's zero spin phase was taken to
   match the midpoint between components at the start of the B-mode of
   that observation or at the beginning of the session if no Q-to-B
   transition was recorded. For L78450, the data from 60 to
   80\,minutes was cut out because of radio frequency interference (RFI).}
 \label{fig:data}
    \end{figure*}

\begin{table*}
\caption{Summary of observations. See Appendix~\ref{subsec:DM} for details of the DM measurements.}             
\label{table:data}      
\centering                          
\begin{tabular}{l c c c c}        
\hline\hline                 
ObsID & Date & Frequency  range,& Obs time per mode,& DM, \\  
      &      & (MHz) & (hours) &  (pc/cm$^3$)\\  
\hline                        
L77924 & 28 Nov 2012 & 25$-$80 &  1.0 B & 15.3348(1)\\
L78450 & 30 Nov 2012 & 110$-$190 & 1.6 B $\rightarrow$ 2.9 Q & taken from L77924 \\
L99010 & 27 Feb 2013 & 25$-$80 &   0.3 Q $\rightarrow$ 3.6 B & 15.3328(1)\\
L102418 & 09 Mar 2013 & 25$-$80 & 2.0 Q $\rightarrow$ 2.0 B & 15.3335(1)\\
L169237 & 21 Aug 2013 & 25$-$80 &  2.0 B $\rightarrow$ 0.5 Q $\rightarrow$ 4.0 B $\rightarrow$ 1.4 Q & 15.3356(1)\\
\hline                                   
\end{tabular}
\end{table*}

B0943 switches between B and Q-mode once in several hours \citep{Suleymanova1984}.
In both modes the average profile consists of two Gaussian-like
components which merge together towards higher frequencies.  The modes
differ in component separation, width and ratio of the component peak
amplitudes \citep{Suleymanova1998}. Such double-peaked profiles are traditionally explained by
a hollow cone emission model \citep{Ruderman1975, Rankin1983,
  Rankin1993}.  According to the radius-to-frequency mapping (RFM)
model, the emission at a certain frequency originates at a fixed
altitude above the stellar surface, with lower frequencies coming from
higher in the magnetosphere \citep{Cordes1978}.  The emitting cone
follows the diverging dipolar magnetic field lines, thus the opening angle of
the cone is larger at lower frequencies and the profile components are
further apart.  \cite{Deshpande2000} found that for B0943 our line-of-sight (LOS) cuts
the emitting cone almost tangentially.  In other words, the angle
between the LOS and magnetic axis $\beta$ is comparable to the opening
angle of the cone $\rho$, with $|\beta/\rho|$ being at least 0.86 (in
B-mode at 25\,MHz). Above 300\,MHz the LOS no longer crosses the
radial maximum-power point of the B-mode cone; for Q-mode this should
happen at even lower frequencies. Thus, at frequencies $\gtrsim 300$\,MHz, only the periphery of the cone can be observed.  This
effect of viewing geometry can plausibly explain B0943's
particularly steep spectrum \citep[$S \propto
  \nu^{-2.9}$,][]{Malofeev2000}.


The two emission modes also have very different single-pulse
properties. In the B-mode the emission is characterized by a series of
coherently drifting sub-pulses, whereas in the Q-mode the emission consists of
sporadic bright pulses, apparently chaotically distributed in phase within the
average profile \citep{Suleymanova1998}.

The transition between modes happens nearly instantaneously, on the
time scale of about one spin period \citep{Suleymanova1998}.  Recently it has been
discovered that this rapid change of radio emission is simultaneous
with a mode switch in X-rays \citep{Hermsen2013}. Using data from the Low-Frequency Array
(LOFAR, 120$-$160\,MHz), 
the Giant Meterwave Radio Telescope (320\,MHz) and XMM-Newton
(0.2$-$10\,keV), these authors showed that B0943 emits bright
pulsed thermal X-rays \textit{only} during the radio Q-mode. This
discovery demonstrates that the long-known radio mode switch is in
fact a transformation of observed pulsar emission across a large part of the
electromagnetic spectrum. Since different parts of the spectrum are
produced by different mechanisms, this broadband switch is a
manifestation of some global changes in the magnetosphere. These
changes could, among other possibilities, be caused by some unknown, non-linear processes
or an adjustment to the magnetospheric configuration
\citep{Timokhin2010}.

The mode switching phenomenon, as well as its relation to other variable 
emission behaviour like nulling, is still far from well understood. Thus 
discovering new features in the portrait of mode-dependent pulsar emission can be helpful
for deciphering the underlying physical processes. Very low radio 
frequencies provide an interesting diagnostic for studying B0943, because in both B and Q-mode
the average profile morphology evolves rapidly with frequency below
100\,MHz \citep{Suleymanova1998}; exploring this evolution may provide 
important clues about the shape and location of the emission regions in each mode.
B0943 has been previously studied at very low frequencies with the
Pushchino and Gauribidanur observatories \citep{Suleymanova1998,Asgekar2001}.  
For both telescopes the observing bandwidth was less than 1\,MHz and a single observing session was no
longer than several minutes -- i.e. not long enough to easily catch mode 
transitions or to follow the evolution of the emission characteristics 
during a multi-hour mode instance.

In contrast, LOFAR enables instantaneous frequency coverage from typically 
either 10$-$90\,MHz or 110$-$190\,MHz, and can track B0943 for up to 6 hours.  
The tremendous increase in bandwidth together with high sensitivity allow
precise measurements of the profile evolution even in the faint Q-mode.
Also, owing to the multi-hour observing sessions, for the first time it becomes
possible to explore the gradual changes in the profile shape within a
single mode instance (at these low frequencies). This gives us important information about
dynamic changes in the pulsar magnetosphere within a single mode and
can be a clue to the mode switching mystery itself.


\section{Observations and timing}
\label{sec:obs_tim}

\begin{figure}
 \centering
 \includegraphics{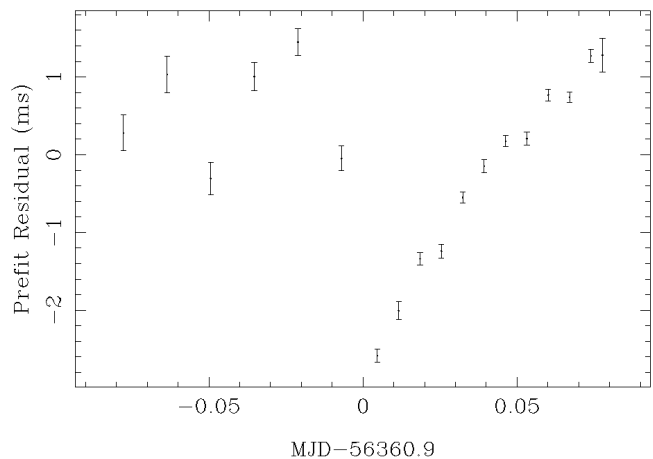}
\caption{Timing residuals from one full LBA observing session (L102418). TOAs from both Q and B modes are shown. The
 transition between Q and B mode happens in the middle of the observation.  }
 \label{fig:timing_comb}
\end{figure}

\begin{figure}
 \centering
 \includegraphics[scale=0.9]{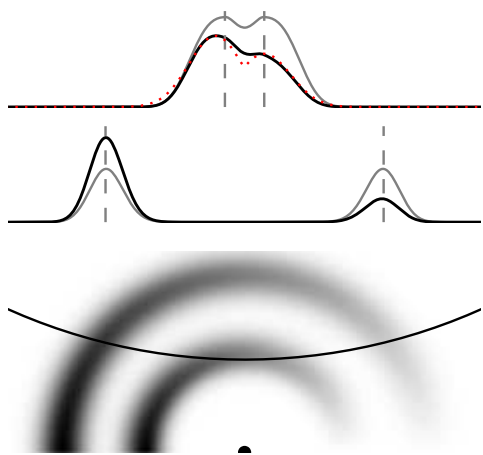}
 \caption{
 A toy model of the average
profile from two cone-like emission regions, centered on the magnetic
axis. For the smaller cone, the LOS does not cut through the inner
side of the cone. Gray profiles correspond to a uniformly illuminated
cone (not shown); black profiles result from the cones with smooth
variation of intensity with magnetic azimuth.  The dotted line shows
the merging component fit model. Vertical dashed lines mark the peaks of 
the uniformly illuminated cone profiles. On the upper subplot, note the slight change in apparent 
position of the leading component for the varying radial intensity case.}
 \label{fig:mc_expl}
\end{figure}

Depending on availability, we observed B0943 with either 19 or 21 LOFAR core
stations.  The signals from the stations were coherently added and the total 
intensity samples were recorded in a filterbank format (see
\citealt{vanHaarlem2013} for an explanation of the telescope
configurations and \citealt{Stappers2011} for a detailed description
of LOFAR's pulsar observing modes). There were five observing sessions
with a total duration of about 20 hours.  Four observations were
conducted with the low-band antennas (LBAs), at a centre frequency of
53.8\,MHz, and with 25600 channels across a 78.1-MHz bandwidth.  One
observation was taken with the high-band antennas (HBAs): 152.24\,MHz
centre frequency, and 7808 channels over 95.3\,MHz bandwidth. The
time resolution of the raw data was 0.65\,ms.  The data were
pre-processed with the standard LOFAR pulsar pipeline \citep{Stappers2011}, resulting in
folded single-pulse archives with 512 phase bins per period and 512
channels per band.  In this paper we describe uncalibrated total
intensity, since flux/polarisation calibration was not fully developed
at the time of data processing.

With the LBAs the pulsar was detected across the range of 25$-$80\,MHz. With the
HBAs, the pulsar was seen from 110$-$190\,MHz, with 135$-$140\,MHz cut
out due to RFI. The one-minute frequency-integrated sub-integrations for
each session are plotted in Fig.~\ref{fig:data} and the mode transitions are
marked with white ticks. 

Data were folded with the ephemerides from \citet{Shabanova2013}. Dedispersion
was done with a more up-to-date and precise 
value of the dispersion measure (DM), obtained by measuring the $\nu^{-2}$ lag of the
fiducial point in the B-mode profile (see Appendix~\ref{subsec:DM} for
details).
In the work of \citet{Shabanova2013}, B0943 was routinely observed during
1982$-$2012 at the Pushchino Observatory at frequencies close to 100\,MHz. 
During this period, B0943 exhibited prominent timing noise, with a characteristic 
amplitude of 220 ms on a timescale of roughly 1500 days. Comparing the timing residuals
from our LBA B-mode observations, we found that our residuals 
are in good agreement with the extrapolation of the timing noise curve from \citet{Shabanova2013},
as they exhibited a long-term linear trend with a net pulse delay of 40\,ms over 270 days.

 \begin{figure*}[t]
   \centering
 \includegraphics[scale=0.8]{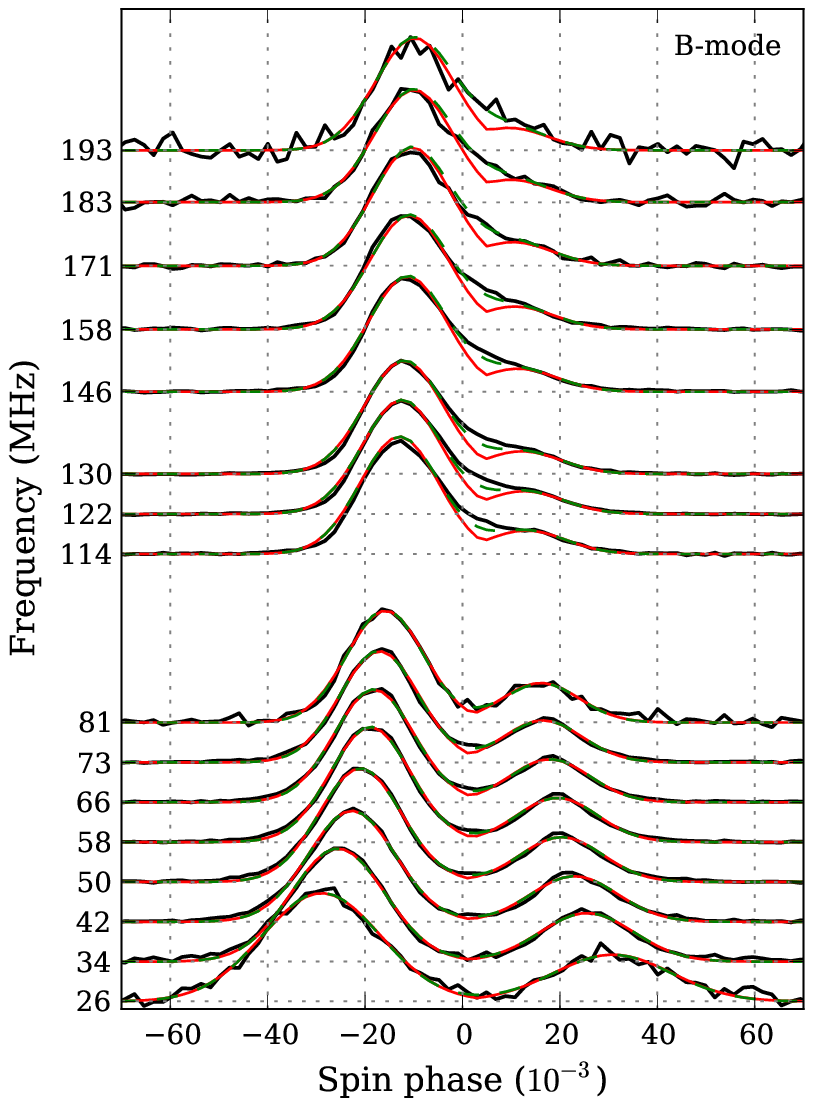}\includegraphics[scale=0.8]{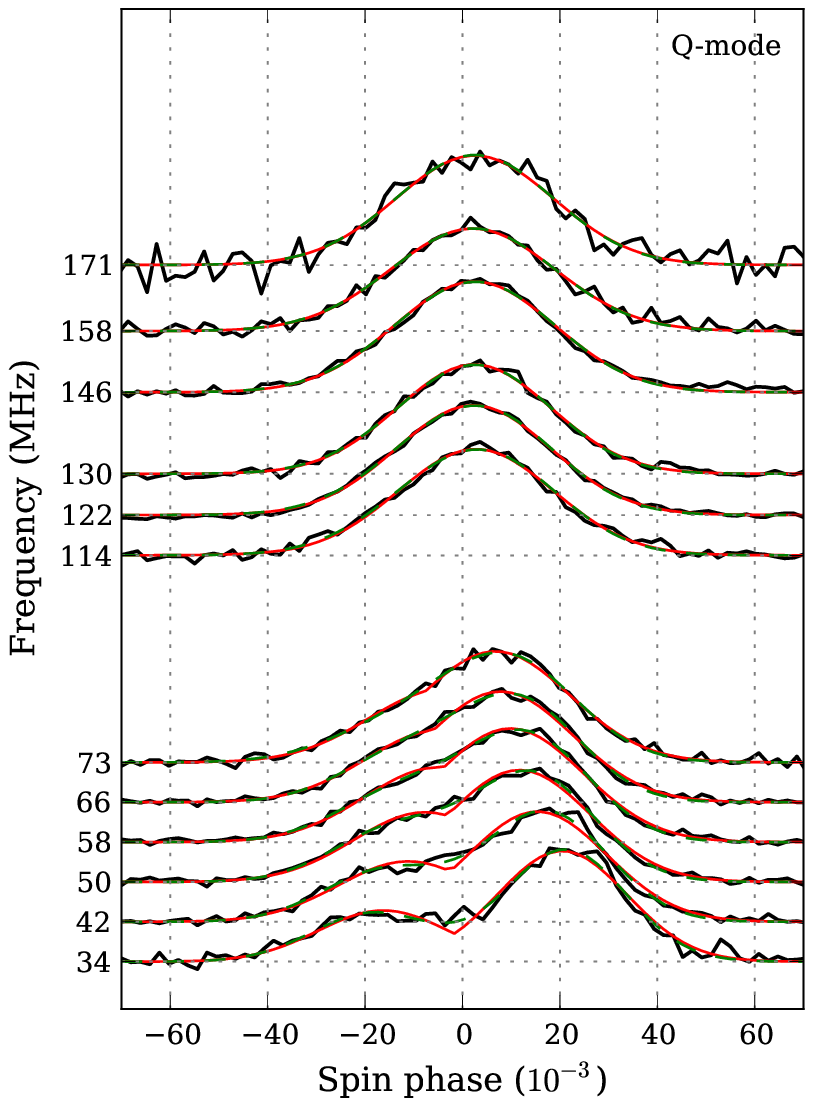}
 \caption{
Frequency evolution of the average profile in B and Q-mode
from observations L102418 in LBA and L78450 in HBA.  
The dashed line (green in the online version) shows the fit from
the OC model (overlapping Gaussians) and the grey line (red in online version) is the MC model
(merging Gaussians).  For the LBA data within each model, both B and Q-mode
were fit with two independent Gaussian components. For the HBA data, the
separation between components, their width and ratio of the peaks were
extrapolated from the frequency dependence within the LBA band and only
the midpoint between components and S/N of the leading component were fit. LBA and HBA
observations are aligned with respect to the midpoint between components in the OC model.}
 \label{fig:modelsQB}
    \end{figure*}

We also found that \textit{within} each observing session the times of arrival (TOAs) of the B-mode pulses 
exhibited a systematic delay with the time
from the start of the mode. This delay (2~ms/hr) is much larger than the expected 
lag due to the long-term timing noise ($\sim$0.006~ms/hr) and is present in the B-mode \textit{only}:  
the residuals from the pulsar's  Q-mode did not show any systematics (Fig.~\ref{fig:timing_comb}).
In Fig.~\ref{fig:timing_comb}, residuals from the observation L102418 are shown. During this observation
B0943 was in the Q-mode for the first half of the session and in the B-mode for the second half.
TOAs were generated using separate templates for the band-integrated B and the Q-mode data.
The templates used are derived from analytic fits using von Mises functions and the PSRCHIVE
task \texttt{paas}\footnote{\texttt{http://psrchive.sourceforge.net/}}. In the B and Q-mode templates the
width at 10\% of the profile maximum was the same, and so we aligned the templates so that they overlap
at this 10\% width. As can be seen there is a significant offset in the arrival time of the first 
B-mode TOA and this offset decreases  exponentially with time until the B-mode TOAs 
essentially align with those in the Q-mode\footnote{As pointed out by the referee, a 
similar result has recently been obtained using many short snapshot observations taken 
as part of the long-term timing campaign on B0943 at the Pushchino observatory (Suleymanova \& Rodin, in prep).}. 
This trend in the B-mode residuals can potentially be explained by a gradual change in profile shape with time:
when cross-correlated with a fixed-shape template, it would result in a growing bias in 
the TOA determination (a similar bias, but for the frequency-dependent
profile evolution, is investigated in \citealt{Hassall2012}, \citealt{Ahuja2007}).
There is indeed evidence that the shape of B0943's profile  evolves within a B-mode:
as was shown by \citet{Suleymanova2009}, the ratio of component peaks 
in B-mode decreases with the time from the start of the mode. 

The observed trend in B-mode residuals motivated us to do
a time-resolved analysis of the profile shape and phase, in order to remove profile evolution bias from
TOA measurements. The results of the analysis 
are presented in Sect.~\ref{subsec:tdep} and discussed in Sect.~\ref{subsec:lag_disc}.

 \begin{figure*}[t]
   \centering
 \includegraphics[scale=0.8]{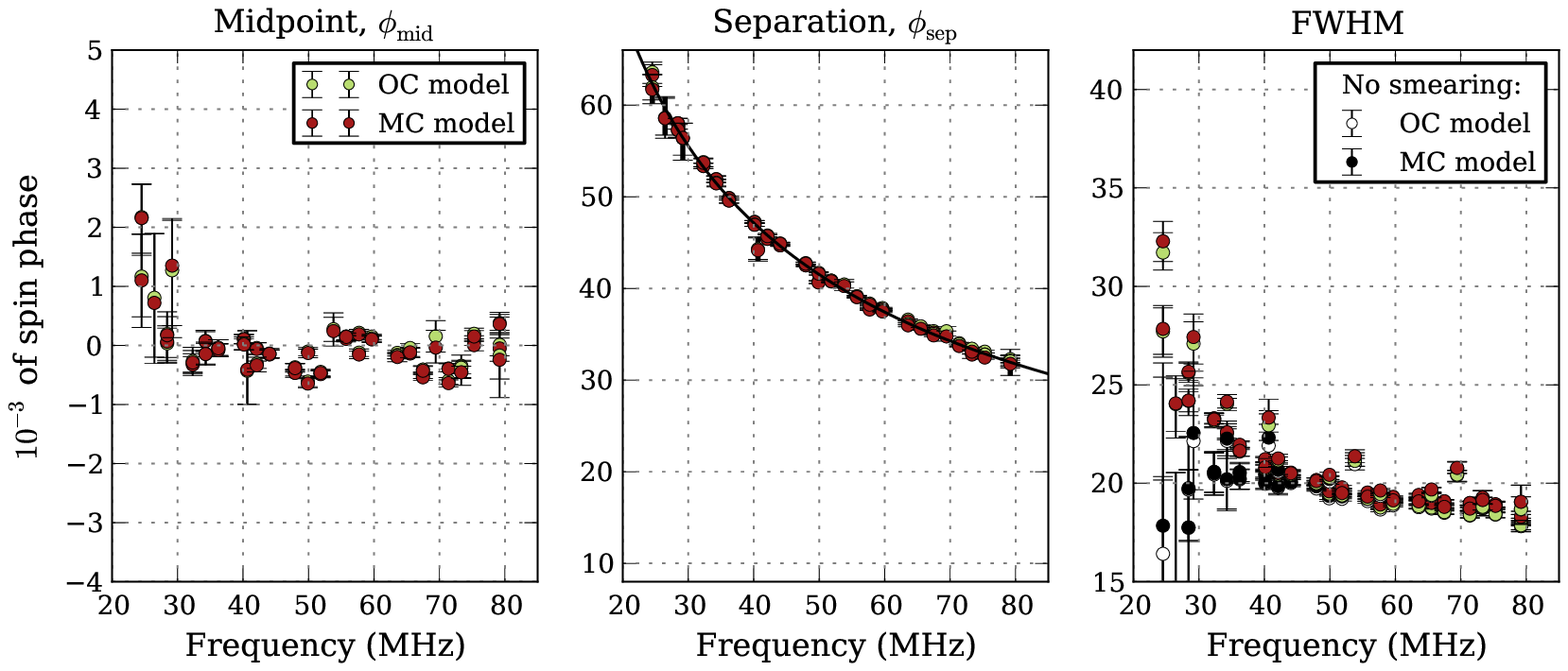}  
 \includegraphics[scale=0.8]{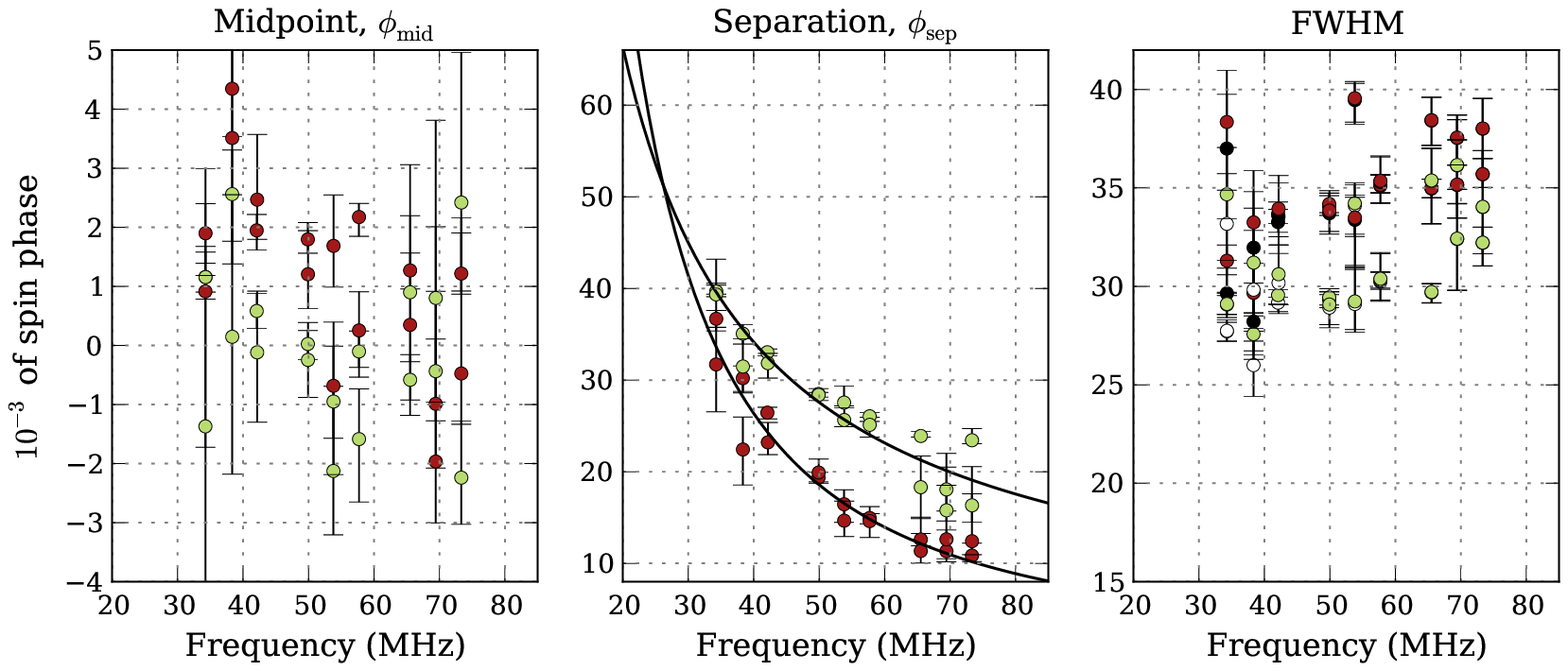}
 \caption{
Frequency evolution of the fitted parameters for OC (lighter circles)
and MC (darker circles) models for all LBA sessions.  The signal was
integrated in time within each session.  \textit{Top row:} B-mode.  For
most frequencies here the MC and OC measurements coincide within the
point markers.  \textit{Bottom row:} Q-mode.  \textit{Left}: the
midpoint between components $\phi_{\mathrm{mid}}(\nu)$.  Phase 0
corresponds to frequency-averaged $\phi_{\mathrm{mid}}$ for the OC model
for each session.  \textit{Middle:} The separation between components
$\phi_{\mathrm{sep}}$, fit with a power-law.  \textit{Right:} the FWHM
of the components. Black and white markers indicate the FWHM corrected 
for intra-channel dispersive
smearing (see text for details).}
 \label{fig:stat}
 \end{figure*}

\section{Modeling the average profile}
\label{sec:aver_prof}

Here we give a phenomenological description of B0943's 
average profile. We decompose the profile into two Gaussian
components and analyse the evolution of the shape and phase of the
components with time and frequency. Starting from some sufficiently high frequency 
(about 80\,MHz for the B-mode and less than 30\,MHz for the Q-mode), the 
two components come very close to each other in phase, and the values of the 
fitted parameters become dependent on the assumptions about the shape of the emitting
region in the pulsar's magnetosphere.  In the following we made two such assumptions
and thus fit the data with two separate models. The first model (``overlapping components'', 
or ``OC model'') assumes that the components come from different parts of 
the magnetosphere\footnote{For example, two separate patches located at different
heights above the stellar surface on field lines originating at different 
magnetic latitudes in the polar cap \citep{Karastergiou2007}.}  and treats the profile as  
the sum of the individual components. 
The second model (``merging components'' or ``MC model'') assumes a cone-shaped 
emission region \citep{Rankin1993}. For closely located components, the observer's 
LOS does not cut through the inner part of the cone, thus the inner part does 
not contribute to the profile (Fig~\ref{fig:mc_expl}). We approximated such profiles 
with two Gaussian components truncated on their inner sides. It should be noted that 
our MC model does not perfectly reproduce the profile from a cone-shaped emission 
region, as it slightly underestimates the intensity between closely located components (Fig~\ref{fig:mc_expl}, dotted line). 
Nonetheless, we include the results from the MC model fitting. They serve as a check on how much 
the fitted model parameters are influenced by the OC model's assumption that
profile components are completely independent. In contrast,
direct modelling of the average profile as 
an emission cone would require accounting for the unknown variation of intensity 
with magnetic azimuth, in order to model components at different emission
heights (Fig~\ref{fig:mc_expl}). 
That approach is not attempted here.

For the profile analysis, we divided the data into several subbands in
frequency ($\nu$) and several sub-integrations in time ($t$) and fit
the average profile in each $(\nu, t)$ cell independently.  The data
from the B and Q-mode were analysed separately\footnote{The high sensitivity of 
the
data allowed us to clearly distinguish the individual pulses. 
The mode transitions were visually identified by the abrupt (within one period) 
start or cessation of the subpulse drifting patterns.}, with the few-second
region around the mode switch being removed. The sub-integrations were
at least 10 minutes long and the subbands at least 5\,MHz wide. The
number of subbands / sub-integrations was reduced if the profile did
not have a sufficient signal-to-noise ratio (S/N$\sim$10).  In each $(\nu, t)$ cell
the $I(\phi)$ profile was fit with two Gaussians using the following
parametrization:

\begin{equation}
I_i(\phi) = I_{c,i} \exp\left[\frac{(\phi-\phi_{c,i})^2}{2\sigma_i^2}\right].
\end{equation}
Here $i=1,2$ for the first (earlier in phase) and second component, respectively. $I_c$
is the amplitude of the Gaussian and $\phi_c$ is the phase of its
centre (in rotational phase). The full width at half maximum (FWHM) of
the component is proportional to $\sigma$:
$\mathrm{FWHM}\approx2.355\sigma$.  A preliminary fit to well-separated
components (B-mode, $\nu<80$\,MHz) showed that the widths of both
components are consistent with being identical, $\sigma_1=\sigma_2$.
In what follows, we assumed that this equality extends for both modes
at all frequencies and fitted both components with the same width
$\sigma$ (although still dependent on $\nu$ and $t$).  Allowing these
widths to be different gives the same quality of fit, but with larger
uncertainty in fitted parameters due to covariance between them.

For the OC model the resulting profile was:
\begin{equation}
I(\phi) = I_1(\phi) + I_2(\phi).
\end{equation}
For the MC model:
\begin{equation}
 I(\phi) = \left[
  \begin{array}{l l}
    I_1({\phi}) & \quad \text{if $I_1>I_2$ }\\
    I_2({\phi}) & \quad \text{if $I_1<I_2$ }
 \end{array} \right.
\end{equation}

For both B and Q-mode the profiles below 100\,MHz are well described
by both OC and MC models (Fig.~\ref{fig:modelsQB}). For both modes and models, 
the simple extension of the frequency dependent LBA fit results
provides a reasonable match to the HBA data (Fig.~\ref{fig:modelsQB}). 
However, the direct fitting of HBA profiles with the OC and the MC models posed some difficulties.
Above 100\,MHz the components of the Q-mode profiles are almost completely merged, 
thus making the two-component fit meaningless.
For the B-mode profiles, the HBA fit parameters appeared
to be inconsistent with an extrapolation from the LBA data, both for the parameter
values and their frequency dependence within the band. 
This may indicate that our simple MC and OC models can not be used to 
describe profiles with closely located components (see Appendix~\ref{subsec:fit} for details).
Thus, in this work we will not cite or discuss the fitted parameters for the HBA
 data, deferring the analysis to subsequent work with a better profile 
model.

\begin{figure*}
  \centering
  \includegraphics[scale=0.9]{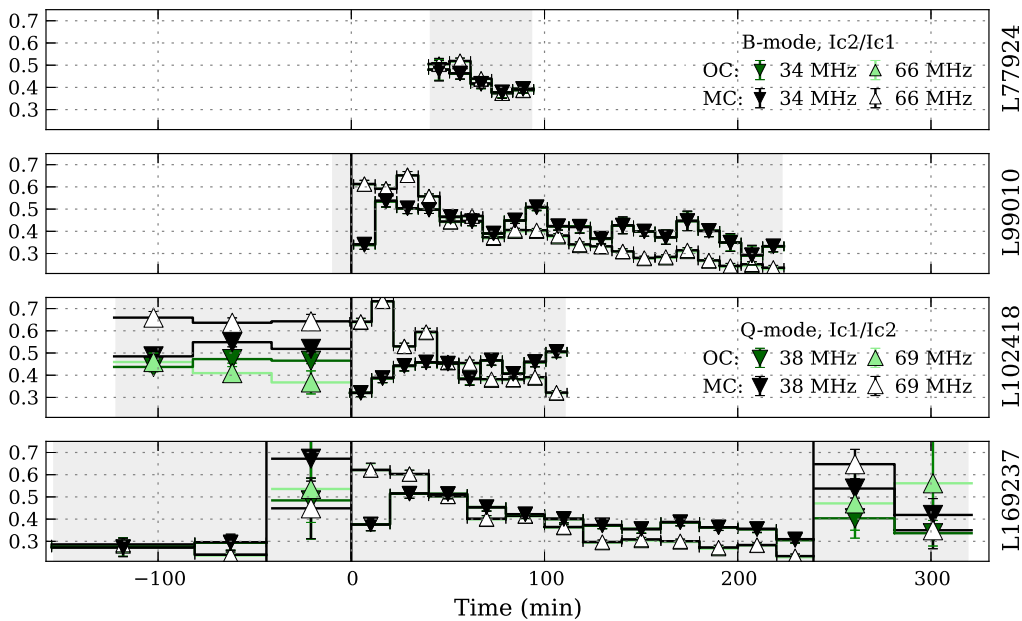}
 \caption{
The ratio of component amplitudes versus frequency and time for all
observing sessions. In order to be able to plot both modes on the same
scale, we give $I_{c2}/I_{c1}$ for B-mode and the inverse quantity,
$I_{c1}/I_{c2}$ for the Q-mode. For the B-mode, $I_{c2}/I_{c1}$
changes with time and this evolution is similar for every mode
instance. This makes it possible to estimate the time since the start of the
B-mode for L77924 (top row), although we did not directly record the
transition in this observation. The vertical black lines mark the transitions between the modes.}
 \label{fig:irat}
 \end{figure*}

\subsection{Spectral evolution of fit parameters}

Fig.~\ref{fig:stat} shows
the frequency evolution of the fitted parameters, 
using B/Q profiles
integrated over all LBA 
sessions\footnote{Here the folding ephemerides for the B-mode were
  adjusted in such a way that the drift of $\phi_{\mathrm{mid}}(t)$ in the
  B-mode (see Sect.~\ref{subsec:tdep}) was zero.}.  For both modes
above 30\,MHz, the midpoint between components does not show any
noticeable dependence on frequency. Judging from the leftmost subplots of Fig. ~\ref{fig:stat}, 
if there exist any
frequency-dependent midpoint shifts, they must be less than $1\times10^{-3}$
of the spin phase for the B-mode and $4\times10^{-3}$ for the Q-mode.

For both modes, the separation between components is well described by
a power-law :

\begin{equation}
 \phi_{\mathrm{sep}} \equiv \phi_{c2}-\phi_{c1} = A \nu^{-\eta}
 \label{eq:xsepfit}
\end{equation}
For both OC and MC models, the power-law index in the B-mode, $\eta\approx0.57(1)$ agrees with
$\eta=0.63(5)$, reported by \citet{Suleymanova1998}. In the Q-mode,
the separation between components has a steeper dependence on
frequency: $\eta\approx1.0(1)$ for the OC and $1.6(1)$ for the MC model.
The frequency-dependent separation of the Q-mode profile components had 
not been investigated before, thus no direct comparison can be
made with the results we present here.

\begin{figure*}
\centering
\includegraphics[scale=0.8]{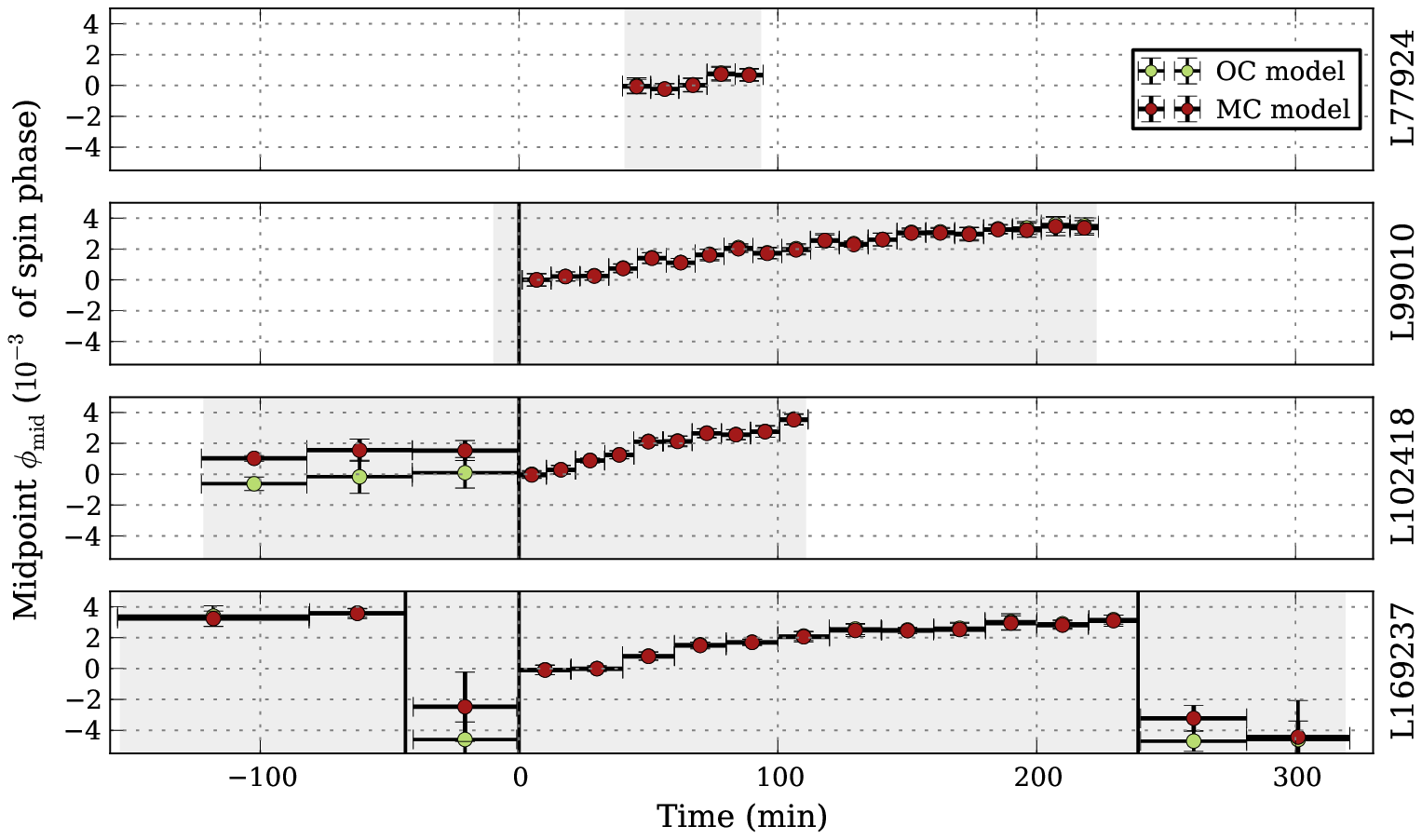}
\includegraphics[scale=0.8]{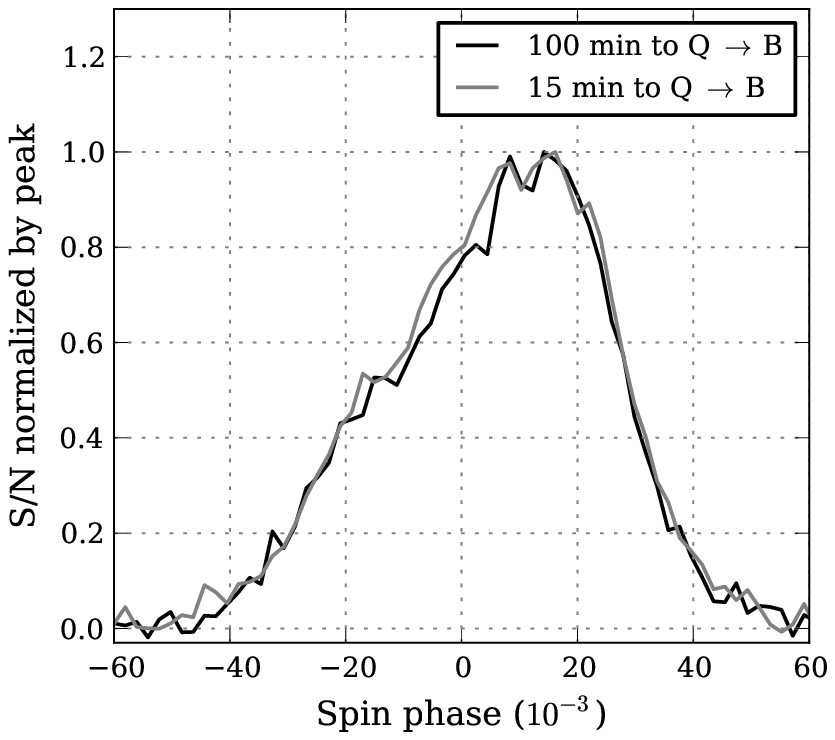}\includegraphics[scale=0.8]{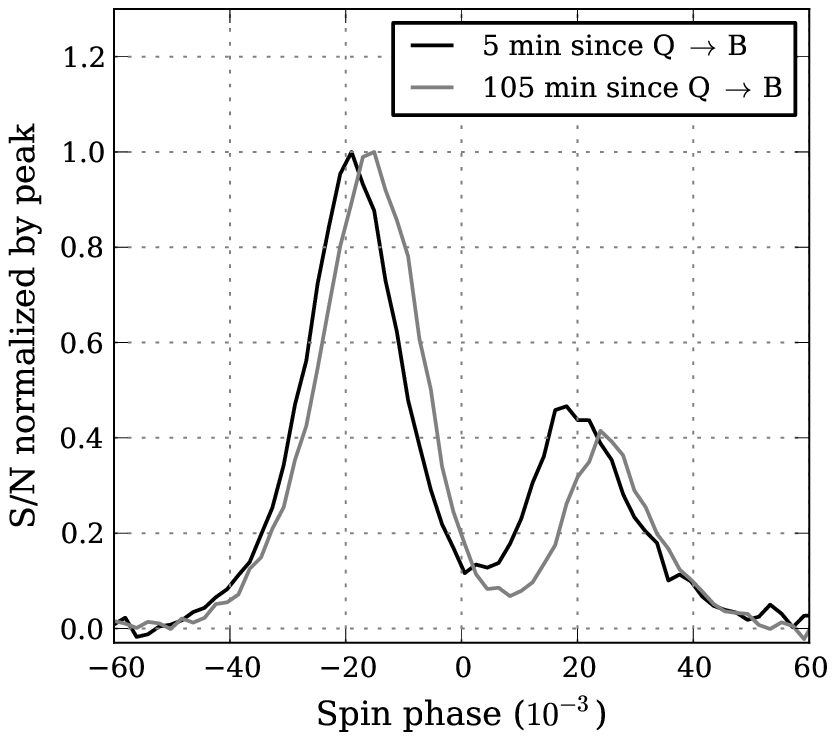}
 \caption{
\textit{Top:} the midpoint between profile components versus time for all LBA
observing sessions.  \textit{Bottom, from left to right:} Two average profiles of the
Q-mode (at the start of observations and just before the mode transition)
and the B-mode (right after mode transition and at the end of the
observation). This is observing session L102418 and the profiles from
Q-mode correspond to the first and third points on the upper plot. For
B-mode, the profiles are from the first and last point of B-mode at
the same subplot of the upper plot. The timing residuals from the same
observations are plotted in Fig.~\ref{fig:timing_comb}. Noticeably, 
in addition to changes in relative intensity of the
components and their width, the profile in B-mode shifts as a whole towards later spin phase.
}
 \label{fig:xmid_vs_t}
    \end{figure*}

The width of the components remains roughly constant above 50\,MHz, with the 
Q-mode FWHM being about 50\% larger than that in B-mode.  Below
50\,MHz the B-mode FWHM grows from $20\times10^{-3}$ of spin phase
to about $30\times10^{-3}$ at 24\,MHz.  However, at these low
frequencies the measured FWHM is substantially affected by our
observing setup, because interstellar dispersion was not compensated 
for within each 2.1-kHz wide channel, introducing additional smearing in time.
To estimate the extra broadening we used Eq.~6.4 from \citet{Lorimer2005}:

\begin{equation}
 \mathrm{FWHM_{meas}} = \sqrt{\mathrm{FWHM_{intr}}^2 + (\delta t/P)^2},
 \label{eq:broad}
\end{equation}
where $\mathrm{FWHM_{meas}}$ is the measured FWHM of the profile, $\mathrm{FWHM_{intr}}$
is the intrinsic FWHM and $\delta t$ is the dispersive broadening within one channel.  Both measured
and intrinsic FWHMs are shown in Fig.~\ref{fig:stat}, in the rightmost
subplots.  Since we are measuring FWHM in frequency bands, averaging the data
from many channels, the resulting FWHM will depend on $\delta t(\nu)$
weighted with an unknown distribution of the uncalibrated peak intensity
within each band.  Two opposite assumptions: that all emission comes
at the upper, or the lower, edge of subband are incorporated into the
errorbars of the intrinsic FWHM values.

The incoherent dedispersion has little effect on the measured FWHMs of
the Q-mode components, since the lower S/N of the data only allows 
measurements  down to 35\,MHz.  For the B-mode, instrumental
broadening can explain most of the rapid growth of FWHM below
30\,MHz. Still, the B-mode components seem to have a slight intrinsic
broadening towards lower frequencies, by about $1\times10^{-3}$ of the
spin phase between 80 and 40\,MHz.

  \begin{figure*}[t]
   \centering
 \includegraphics{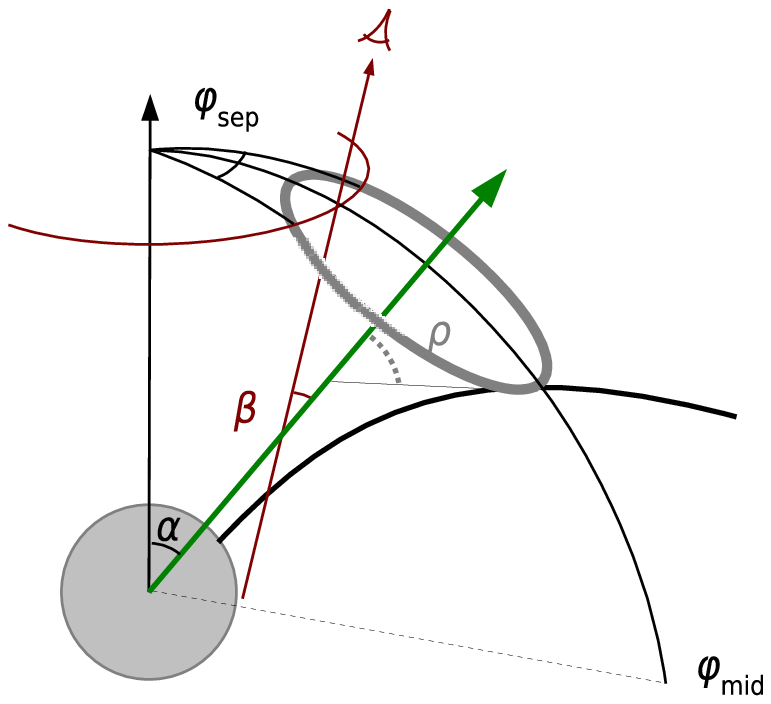}\includegraphics{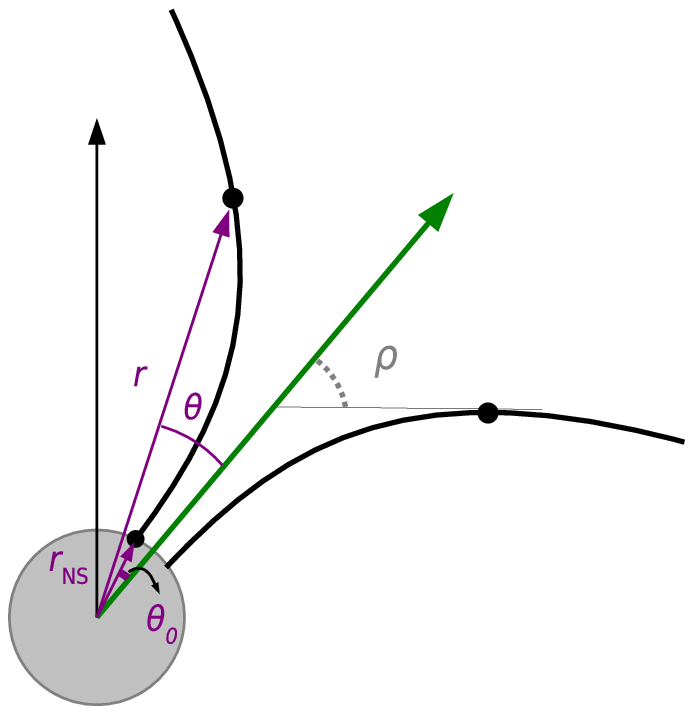}
 \caption{
\textit{Left:} A toy model of the pulsar magnetosphere.  The pulsar's
spin axis is vertical and the magnetic axis is inclined by the angle
$\alpha$. An example  dipolar field line 
is shown by the thick black line. The position of
the LOS vector (red in the online version) stays fixed in space while the pulsar rotates
around its spin axis.  The plot shows the moment when spin, magnetic
and LOS vectors are lying in the same plane (the fiducial plane). At this moment the angle
between observer and magnetic axis is $\beta$. Usually $\beta$ is
measured clockwise from the magnetic axis, so for the vectors here the
value of $\beta$ is negative. Here we assume that emission comes from
a cone at fixed height above the pulsar surface (with respect to
the magnetic axis). The opening angle of the cone is marked with a thick dotted
grey line. All angles in this plot are identical to the ones in \citet{Lorimer2005}, Fig.~3.4a. 
However, in order to make the relationship between the angles more clear, 
in their figure $\beta$ and $\rho$ are plotted as originating 
from the centre of the star. Such transformation does not affect Eq.~\ref{eq:rho},
since $w$ ($\phi_\mathrm{sep}$ in our plot) stays the same for both cases. 
\textit{Right:} The relationship between the opening angle
of the emission cone $\rho$ and coordinates of the emitting point
$(r,\theta)$.
}
 \label{fig:psr_cartoon}
  \end{figure*}

\subsection{Time-resolved evolution of fit parameters}
\label{subsec:tdep}

Earlier work on B0943 revealed slow changes of the B-mode profile
shape during the mode.  These changes were
characterized by $I_{c2}/I_{c1}(t)$ -- the evolution of the ratio of component
peaks with the time since mode onset.  Here we confirm the
result of \citet{Suleymanova2009}, who carried out their analysis at
similar observing frequencies.  According to our observations, 
right after a Q-to-B transition, $I_{c2}/I_{c1}$ strongly depends on observing frequency,
growing from $0.35\pm0.05$ at 34\,MHz to $0.65\pm0.05$ at 66 MHz (Fig.~\ref{fig:irat}). 
This tendency seems to hold true at higher observing frequencies: 
at 112\,MHz the ratio is $1.2\pm0.2$ \citep{Suleymanova2009} and
at 327\,MHz it reaches $1.75\pm0.05$ \citep{Rankin2006}.
During the first 10$-$30~minutes from the start of the B-mode (hereafter ``initial phase''),
the $I_{c2}/I_{c1}$ below 100\,MHz increases to about 0.5$-$0.6 (our data) and decreases
above 100\,MHz to $1\pm0.2$ \citep{Rankin2006,Suleymanova2009}. The duration of the initial phase seems to be smaller
at higher frequencies (less than 10 min at 327\,MHz and up to 30 min
at 34\,MHz).  After the initial phase, $I_{c2}/I_{c1}$ starts gradually decreasing at all 
observing frequencies, reaching $0.3\pm0.1$ at 34\,MHz (our data) and $0.2\pm0.05$ at 327\,MHz \citep{Rankin2006}
by the next B-to-Q transition.
Such behaviour of $I_{c2}/I_{c1}$ was similar for all B-mode
instances and we used this fact to estimate the age (time from the
start) of the B-mode for observing session L77924, where the Q-to-B
transition happened before the observing session started. At the start
of the observations $I_{c2}/I_{c1}$ at 34 and 66\,MHz was the same and
equal to 0.5, which corresponds to about 40\,min from the start of the
mode.

In the Q-mode the ratio of component peaks did not have any
noticeable change with time. However, in the Q-mode the S/N is much
lower, thus the ratio cannot be constrained as well as in the
B-mode.  The same applies to the FWHM: in Q-mode it did not have any
temporal dependence. In the B-mode, the FWHM (not corrected for instrumental
broadening) grew with the age of the mode: from $20\times10^{-3}$ to
$25\times10^{-3}$ of the spin phase at 38\,MHz (with typical error of measurement of 
$0.4\times10^{-3}$) and from
$19\times10^{-3}$ to $20\times10^{-3}$ at 69\,MHz (with typical error of measurement of 
$0.2\times10^{-3}$).

As was suggested by the observed changes in arrival time discussed in Section~\ref{sec:obs_tim}
and visible in Fig.~\ref{fig:timing_comb}, we also see a change in the B-mode profile midpoint 
as a function of time since the onset of each B-mode. 
In Fig.~\ref{fig:xmid_vs_t}, the profile midpoint $\phi_{\mathrm{mid}}(t)$ was
obtained by weighted averaging of $\phi_{\mathrm{mid}}(\nu,t)$ in every
($\nu,t$) cell.  The error is the standard deviation of the weighted
sample, about $0.3\times10^{-3}$ of the spin phase.  
In B-mode the midpoint gradually moved towards later spin
phases with an average rate of 2\,ms ($2\times10^{-3}$ of spin phase)
per hour.  The shift was more rapid at the start of the B-mode and
flattened out with time. 
In the Q-mode $\phi_{\mathrm{mid}}(t)$ was constant with time within the
error of a single measurement ($2\times10^{-3}$ of spin phase for the longest and
highest S/N Q-mode observation, L102418).  For both the OC and MC
models, and for both of the observed Q-to-B transitions, the Q-mode
midpoint seems to coincide with the initial B-mode midpoint within at most
 4.5 errors of measurement.  Based on the L169237 observation, the
Q-mode midpoint is the same in every instance of the Q-mode, although
more observations are needed before confirming this with certainty.

The size and rate of change of the B-mode midpoint location offset 
(Fig.~\ref{fig:xmid_vs_t}) exactly matches that seen for the TOAs (Fig.~\ref{fig:timing_comb}).
For the midpoint location, we also see a clear phase step at the transition between the
B and Q-mode  corresponding to about 4 milliseconds. Due to the
TOAs and the profile variation models having different fiducial points
in the two modes, this is seen as a step from B to Q and Q to B in
the TOAs and profile variation models respectively. 

We did not find any changes in separation between components
throughout any of the B-modes we observed. At any given frequency,
$\phi_\mathrm{sep}(t)$ measurements did not have an obvious dependence on time and
for the sessions with best S/N they had standard deviation of $0.4\times10^{-3}$
of the spin phase.

\section{Discussion}
\label{sec:disc}

Here we give one possible interpretation of the frequency and time
evolution of the average profile. We assume a purely dipolar field,
with the magnetic axis inclined with respect to the spin axis by the
angle obtained in \citet{Deshpande2000}. We assume RFM, however we do
not require the emission region to be cone-shaped. Our reasoning
thus 
also applies to models where radio emission can come from
multiple 
patches around field lines starting at different magnetic
latitudes, 
as long as RFM holds within every single patch \citep{Lyne1988,Karastergiou2007}.

\subsection{Verifying the basic geometry}

The derivation of B0943's geometry done by \citet[hereafter
  DR2001]{Deshpande2000} was partially based on the frequency
evolution of the width of the average profile. In this section we will
examine if our measurements of average profile width agree with DR2001.

First, we review the derivation done by DR2001. The notation for
angles
 is shown in Fig.~\ref{fig:psr_cartoon}. In the Figure,
the case for negative $\beta$ (LOS passes between spin and magnetic
axis) is shown.  This is called ``inside traverse''.  If $\beta>0$,
the LOS would be on the other side of the magnetic axis and this is an ``outside
traverse''.  The relationship between profile width $w(\nu)$ and
opening angle of the emission cone at that frequency can be established
with spherical geometry \citep{Gil1984}:

\begin{equation}
 \cos(\rho) = \cos(\alpha)\cos(\alpha+\beta) + \sin(\alpha)\sin(\alpha+\beta)\cos(w/2)
 \label{eq:rho}
\end{equation}
DR2001 define $w$ as the distance between the two half-maximum points in
B-mode.  Then, they assume that $\rho(\nu)$
obeys the empirical relation from \citet{Thorsett1991}, given here in the
form adopted in DR2001:

\begin{equation}
\rho(\nu) = \rho_{1\,\mathrm{GHz}}(1 + 0.066\cdot\nu_{\mathrm{GHz}}^{-a})P_{\mathrm{s}}^{-0.5}/1.066
\label{eq:rho_nu}
\end{equation}
\citet{Mitra1999} analysed 37 non-recycled pulsars with conal profiles and
showed that their opening angles at 1\,GHz, $\rho_{1\,\mathrm{GHz}}$,
tend to group around the values of $4.3^{\circ}$, $4.5^{\circ}$ and $5.7^{\circ}$,
although the scatter within these groupings is comparable to their separation. 
The smallest and largest values are called ``inner''
and ``outer'' cones, after \citet{Rankin1993}.

DR2001 use the observed $w(\nu)$ together with Eq.~\ref{eq:rho} and
Eq.~\ref{eq:rho_nu} to fit for $a$, $\alpha$ and $\beta$.  They fixed
$\rho_{1\,\mathrm{GHz}}$ at the values for the inner and outer cone. This
still left room for several solutions. The possibilities were narrowed down 
 by drifting subpulse analysis, namely matching the predicted
values for the longitude interval between drifting subpulses 
$P_2$ and the rotation of position angle between subpulses.
Only the inner traverse ($\beta$<0) satisfied the observed data. In the case of an
 inner cone $\alpha=11.58^\circ$, $\beta=-4.29^\circ$ and for an
outer cone, $\alpha=15.39^\circ$, $\beta=-5.69^\circ$.  These two
models have identical predictions, and, as the authors note, any
reasonable value of $\rho_{1\,\mathrm{GHz}}$ would behave similarly. That means
that we do not really know $\alpha$ and $\beta$, but since most
pulsars fall between inner and outer cones we can take these two
sets of values as boundaries.

\begin{figure}
   \centering
 \includegraphics[scale=0.8]{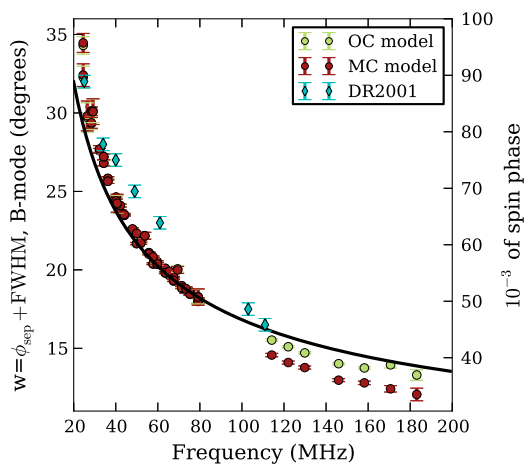}
 \caption{
The width of the average profile in the B-mode between two half-maximum
points, $w\equiv \phi_{\mathrm{sep}(\nu)}+\mathrm{FWHM}(\nu)$.  The
cyan diamonds are the measurements of profile width from
\citet{Deshpande2000}. The green and red circles are from our data
(both LBA and HBA), fitted with the OC and MC models, respectively. 
At frequencies below 40\,MHz the profile width is somewhat 
affected by intra-channel
dispersive smearing. The most simple model is shown by the black line: it
assumes $\phi_{\mathrm{sep}}$ from the LBA fit and the FWHM derived from the LBA data
above 40\,MHz.
}
 \label{fig:w_nu}
  \end{figure}
    
In our notation, $w(\nu)\equiv
\phi_{\mathrm{sep}}(\nu)+\mathrm{FWHM}(\nu)$.  Our values for $w(\nu)$
fall close to DR2001 (Fig.~\ref{fig:w_nu}), with the discrepancy
potentially being explained by the small fraction of the Q-mode data in their folded profiles \citep{Deshpande2000}. 
In Fig.~\ref{fig:w_nu} we also plot $w(\nu)$ for the independent fit in the HBA data. Although
$\phi_{\mathrm{sep}}$ and FWHM are covariant and do not match the
extrapolated values from the LBA data (see Appendix~\ref{subsec:fit}), their sum represents the actual 
width of the profile, which does not suffer
so much from the complications of 
determining the individual phases
and widths of the components.  The small difference
between the two models in the HBA data is due to discrepancy in estimating the peak S/N
of the components, which affected the S/N of half-maximum and thus the
value of FWHM\footnote{For the HBA data, profile components are so close to each other that
in the OC model each component has a significant level of emission at the phase of the other component's centre. 
Thus, the fitted S/N of the profile peak is smaller than in the MC model, 
which results in the lower half-maximum threshold and the larger fitted FWHM.}. 
Below 40\,MHz the component widths are significantly affected by
intra-channel 
dispersive smearing, although some hint that the components also broaden intrinsically does exist (see
Fig~\ref{fig:stat}). For comparison, in Fig.~\ref{fig:w_nu} we plot
$w_\mathrm{extrap}(\nu)=0.384\nu^{-0.567} + 0.0185$, with the 
 FWHM and component separation from the B-mode LBA data at 40$-$80\,MHz.

Unfortunately, we cannot fully repeat the DR2001 derivation because it
relies on polarimetric information, which we do not have. However, we checked if
 $\rho(\nu)$ determined from our data satisfies Eq.~\ref{eq:rho_nu} at a given geometry. We
converted $w(\nu)$ to $\rho(\nu)$ using Eq.~\ref{eq:rho} for the two possible sets
of ($\alpha$, $\beta$) and then fitted it with Eq.~\ref{eq:rho_nu}. We excluded
$\nu<40$\,MHz data from the fit due to intra-channel dispersive broadening. For both
inner and outer cone geometries the widths from the OC/MC model can
be reasonably fit with
Eq.~\ref{eq:rho_nu}, with $\rho_{1\,\mathrm{GHz}}$ lying close to the
implied values: $6.1^{\circ}$~/~$6.0^{\circ}$ for the
outer cone and $4.6^{\circ}$~/~$4.5^{\circ}$ for the
inner cone geometries. The power-law index $a$ was $-0.55$~/~$-0.58$,
somewhat higher than $-0.396$ from DR2001.

In summary,  our measured $w(\nu)$ agrees with both the inner and the outer cone geometries from
the DR2001 if one assumes a steeper frequency dependence of the profile width.  
Further discrimination between the two sets of angles will require
polarimetric data.

\subsection{Where does the emission originate?}
\label{subsec:emr}

B0943's spin period of 1.098\,s places its light cylinder at
$r_{\mathrm{LC}} = Pc/(2\pi)=5.24\times10^4$\,km.  The magnetic
latitude of the last open field line is
$\theta_p=\arcsin(\sqrt{R_{\mathrm{NS}}/R_{\mathrm{LC}}})=0.791^{\circ}=47.5'$,
assuming $R_{\mathrm{NS}}=10$\,km.
 
In the previous section, Eq.~\ref{eq:rho} was used to map the profile
width to the opening angle of the emission cone. Note that this
equation does not necessarily require the emission region to be
cone-shaped; it simply sets the relation between a) the distance from a
chosen profile component to the fiducial point and b) the angle
between the corresponding emitting patch and magnetic axis. We use 
`fiducial point' here to
describe the frequency-independent spin phase when the LOS
passes the plane defined by magnetic and spin axes.  In principle,
emitting patches can be completely independent and components can have
different $\phi_{\mathrm{sep}}(\nu)$ with respect to the fiducial
point, as was indeed observed for PSR~B0809+74 by \citet{Hassall2012}.

For B0943, the components evolve symmetrically around a
frequency-independent midpoint, making the latter an ideal candidate
for the fiducial point. In this section we adopt the assumption that
the emission region has circular symmetry around the magnetic axis and
will refer to it as a cone, although all results obtained will also be
valid for independent patches.

\begin{figure*}[t]
   \centering
 \includegraphics[scale=0.8]{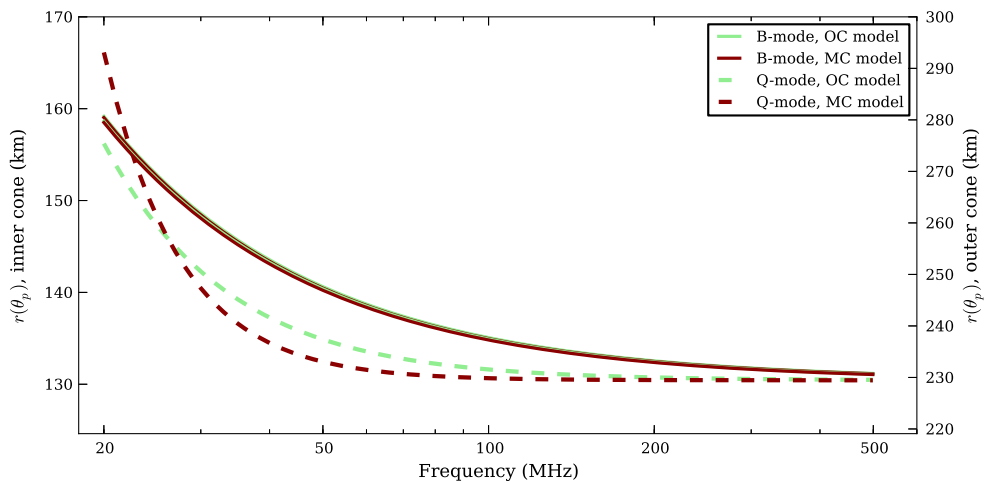}
 \caption{Emission heights (radial distance from the centre of the star) 
 for a cone originating at the last open field line (magnetic latitude $\theta_p$). 
 For any other starting magnetic latitude $\theta_0$, $r(\theta_0)\approx r(\theta_p)\times(\theta_p/\theta_0)^2$.  
 The emission heights at frequencies above the LBA band were extrapolated from 
 the parameters derived from the LBA data. 
}
 \label{fig:em_rads}
  \end{figure*}

The opening angle of the emission cone (found from Eq.~\ref{eq:rho}
with $w =\phi_{\mathrm{sep}}$) can be related to the polar coordinates
$(r, \theta)$ of the field line (Fig.~\ref{fig:psr_cartoon},
right). If the emission region is close to the magnetic axis
($\theta\lesssim20^\circ$, $\rho\lesssim30^\circ$, true for all our
data), then $\theta\approx2\rho/3$. Combining this with the equation for a
dipolar field line:

\begin{equation}
 \frac{\sin^2\theta}{r}=\frac{\sin^2\theta_0}{R_\mathrm{NS}},
 \label{eq:fieldline}
\end{equation}
we can obtain the coordinates of the centres of the emitting patches in
the pulsar magnetosphere, $(r(\nu), \theta(\nu))$. However, for each $\nu$
the coordinates are not unique, since we do not know the magnetic
latitude of the foot of the field line, $\theta_0$.  The upper limit
on $\theta_0$ (and thus lowest possible emission height $r(\nu)$ for
every $\nu$) comes from the requirement that radio emission originates in
the open-field-line region, thus $\theta_0 < \theta_p$.  The lower
limit on $\theta_0$ can be deduced from the absence of
frequency-dependent arrival delays in the midpoint
$\phi_\mathrm{mid}(\nu)$. For this, we use the derivations of
\citet{Gangadhara2001}. If frequencies $\nu_1$ and $\nu_2$ are emitted
at radii $r_1$ and $r_2$ from the centre of the star, then the light
travel time delay (retardation) between those two frequencies will be:

\begin{equation}
 t(\nu_2)-t(\nu_1) = \frac{r_1}{c}-\frac{r_2}{c} = \frac{P}{2\pi}\left(\frac{r_1}{r_\mathrm{LC}}-\frac{r_2}{r_\mathrm{LC}}\right).
 \label{eq:retard}
\end{equation}

The azimuthal corotation velocity $2\pi r/P$ is bending the emission
beam in the direction of the pulsar rotation (aberration):

\begin{equation}
 t(\nu_2)-t(\nu_1) = \frac{P}{2\pi}\left[ \sin^{-1}\left(\frac{r_1\sin(\alpha+\beta)}{r_\mathrm{LC}}\right)-  \sin^{-1}\left(\frac{r_2\sin(\alpha+\beta)}{r_\mathrm{LC}}\right) \right].
 \label{eq:aberr}
\end{equation}



Comparing the sum of the light travel and aberration delays\footnote{
We did not include the magnetic sweepback delay \citep[proportional to 
$\sin^2\alpha\times((r_2-r_1)/r_\mathrm{LC})^3$, ][]{Shitov1983}, because
for our values of $\alpha$ and estimated emission heights 
it is much smaller than the aberration and retardation delays.} 
with the observed
$\phi_\mathrm{mid}(\nu)$ (Fig.~\ref{fig:stat}, left), we conclude that
the net delay must be less than 1\,ms between 25 and 80\,MHz in B-mode and
at most 4\,ms between 35 and 75\,MHz in Q-mode.  This places the minimum
$\theta_0$ at about $0.1\theta_p$ for the Q-mode and about
$0.27\theta_p$ for B-mode. The exact lower limits on $\theta_0$ for
inner and outer cone and OC/MC models are given in
Table~\ref{table:theta0}.

Fig.~\ref{fig:em_rads} shows the emission heights for the last
open field line, $r(\theta_p)$. For any other $\theta_0$, 
$r(\theta_0)\approx r(\theta_p)\times(\theta_p/\theta_0)^2$.
The heights for the B-mode and inner cone geometry
are in good agreement with \citet{Deshpande2000}.  Note that if
$\phi_\mathrm{sep}$ derived from the LBA data does not change dramatically above
the LBA band, then the emission at basically all frequencies above
20\,MHz comes from quite a narrow range of heights:
13$-$300$R_{\mathrm{NS}}$, depending on the chosen $(\alpha, \beta)$ 
solution and $\theta_0$. At
any given frequency above 20\,MHz, the heights do not differ much
between the modes, with $0.94<r_\mathrm{Q}/r_\mathrm{B}<1.04$, assuming both
modes come from the same $\theta_0$. Emission is much closer to the
surface of the star than to the light cylinder:
$r/r_\mathrm{LC}<0.06$. 

\begin{table}[t]
\caption{Minimum magnetic latitude $\theta_0$}             
\label{table:theta0}      
\centering                          
\begin{tabular}{c c c }        
\hline\hline                 
& Inner cone & Outer cone\\
\hline                        
B-mode & $11'$ & $15'$\\
Q-mode, OC model & $5'$ & $7'$ \\
Q-mode, MC model & $6'$ & $8'$\\
\hline                                   
\end{tabular}
\end{table}

Although the MC model is only an approximation to the profile produced
by a cone-shaped emission region, the
results obtained within this model can still be used to estimate 
the width of the cone.  At the surface of the
star:

\begin{equation}
 \Delta \theta_0 \leq 2\theta_0 \left(\dfrac{\rho(w=\phi_\mathrm{sep}+\mathrm{FWHM})}{\rho(w=\phi_\mathrm{sep})}-1\right).
\end{equation}
The $\leq$ sign reflects the fact that the radial distribution of intensity
within the cone is not symmetric around the maximum point, with
the radial intensity dropping faster at the inner side of the cone\footnote{Note that $\rho(w)$ 
from Eq.~\ref{eq:rho} is not a linear function of $w$. In order to reproduce an observed
Gaussian-shaped component, the inner and outer half-widths of the cone must be different.}. Both allowed
geometries give the same value for $\Delta\theta_0$. For B-mode at
30\,MHz $\Delta\theta_0\leq4'\times\theta_0/\theta_p$, for the Q-mode
$\Delta\theta_0\leq8'\times\theta_0/\theta_p$. At 80\,MHz the width of
the cone is about 2 times smaller.

\subsection{What causes the lag of the B-mode midpoint?}
\label{subsec:lag_disc}
B0943's average profile undergoes noticeable evolution during each
B-mode. Both the ratio of the component
peaks and their widths change 
in a frequency-dependent manner, and the midpoint between the two components,
$\phi_\mathrm{mid}$, systematically shifts towards later spin
phases in the LBA data. The cumulative midpoint lag per mode instance is
$\Delta\phi_\mathrm{mid}\approx4$\,ms, or $4\times10^{-3}$ of the spin
phase. Throughout the B-mode, $\phi_\mathrm{mid}$ is independent of observing
frequency down to the error of measurement of 0.3\,ms. The frequency-dependent
separation between components, $\phi_\mathrm{sep}$, also does not change during
the B-mode: at any given frequency $\phi_\mathrm{sep}$ is constant down to
the typical error of measurement, 0.4\,ms.

Our single HBA observation did not reveal any $\phi_\mathrm{mid}$ drift within
2\,ms, although we only recorded a 100-min part of the B-mode right
before B-to-Q transition. The observed lag rate in the LBA data is not uniform,
decreasing towards the end of the mode (Fig.~\ref{fig:xmid_vs_t}), and
the cumulative lag in the last 100\,min is typically less than 2\,ms.
Also, at higher frequencies, where components are close to each other,
it is important to develop better, more physically adequate 
profile models in order to avoid contamination of midpoint measurements 
by changing peak ratio (see Appendix~\ref{subsec:fit}).

The LBA band Q-mode midpoint is
constant within the uncertainties. However, 
uncertainties in  the Q-mode are a few times larger than in the
B-mode and the Q-mode is shorter, making it harder to detect any Q-mode
lag. From the observing session with the highest S/N in the Q-mode (L102418),
the cumulative Q-mode midpoint lag rate is less than 1\,ms/hr.
Judging by the two available Q-to-B transitions, the midpoint in
Q-mode coincides with the initial B-mode midpoint. On the other hand, at
the B-to-Q (2 recorded) transitions there is a 4$-$6\,ms jump to
earlier rotational phase.  We had only one observing session with two
B-to-Q transitions (L169237) and on that day the midpoint was at the
same phase at the end of both B-mode instances.

In what follows we propose several tentative explanations for the
observed midpoint lag and discuss their plausibility.

\subsubsection{Location of the emission regions} 
\label{ssub:emh}

If the emitting region is gradually moving against the sense of 
the pulsar's rotation (with different $\theta_0$
being illuminated over time), the pulse profile will appear to lag. 
However, in this case $\phi_\mathrm{mid}$ and $\phi_\mathrm{sep}$ will
depend both on the frequency and time from the start of the mode (unless
emission heights are adjusting themselves in a peculiar way that would 
keep $\phi_\mathrm{mid}(\nu)$ and $\phi_\mathrm{sep}(t)$  constant).
The 4\,ms ($86.4'$) cumulative lag of the midpoint at the derived emission heights
corresponds to a $0.57'$ shift of the magnetic latitude of the centre of the 
emission cone at the pulsar's surface. If at the beginning 
of a B-mode the magnetic latitude of the cone centre is $0.29'$ and during the mode the cone
gradually moves through the magnetic pole up to $\theta_0=0.29'$ on the other side of the fiducial
plane, then $\phi_\mathrm{sep}(t)$ will stay 
within the error of measurement and $\phi_\mathrm{mid}(\nu)$ within 2.5 measurement
errors. Here we assumed that the emission heights do not change throughout the mode,
however, some variation of emission heights is not excluded, as long
as the net midpoint and separation stay within the measurement errors. 



\subsubsection{Variation of the cone intensity with magnetic azimuth}

If the components are partially merging (the LOS does not cross the inner cone
boundary), then the measurements of the midpoint position are biased if 
the components have different intensity (Fig.~\ref{fig:mc_expl}, compare the peak phases of
partially merged profiles  for the uniformly illuminated cone
and the cone with the smooth variation of radial intensity with magnetic azimuth). 
The amount of bias depends on the ratio of the component peaks and thus will vary
with time if the peak ratio varies.  However, the direction of the
midpoint shift caused by this effect should be opposite to that
observed here: the midpoint shifts towards the brighter component, the
leading one for the B-mode (which becomes only relatively brighter
with the age of the mode, see Fig.~\ref{fig:irat}). Also, this effect
is frequency-dependent, being virtually zero for the lower frequencies,
where components are well separated.

\subsubsection{Spin-down rate} 
 
\citet{Lyne2010} showed that, for at least some
pulsars, long-term variability 
in the radio profile shape is
correlated with changes in the pulsar spin down rate.
This may be true for B0943 as well, however the
amplitude of this effect is too small to explain the observed midpoint
lag of the B-mode profile.  The delay of 2\,ms/hr corresponds to
$\delta\dot{P}\approx10^{-9}$~s/s, $10^6$ larger than the pulsar's
spin-down rate, $\dot{P}$.  That is, however, many
orders-of-magnitude larger than the fractional changes in $\dot{P}$ found in PSRs B1931+24
(50\%; \citealt{Kramer2006}), J1841$-$0500 (250\%;
\citealt{Camilo2012}) and 1832+0029 (77\%; \citealt{Lorimer2012}).
Furthermore, a spin-down rate change cannot explain the phase jump at
B-to-Q transition. We thus conclude a different mechanism is likely at
play.
 
\subsubsection{Toroidal magnetic field}

In a pulsar magnetosphere, radio waves are emitted in the direction of
the magnetic field, thus any changes to the field lines at the emission
heights will affect the direction of the beam. An observed shift of
fiducial point towards later phases can be interpreted as evidence for a
varying toroidal component of the magnetic field (field lines bending
against the sense of rotation).  A toroidal component can appear from
the interaction between the rotating magnetic field and the frozen-in
plasma.  This effect was investigated by \citet{Michel1969}, who made
a relativistic expansion to the Weber-Davis model of the solar wind. Using
 Eq.~8 from \citet{Michel1969}, the midpoint lag can be written as:
 \begin{equation}
 \Delta\phi_\mathrm{sep}=\frac{\Delta B_\phi}{B_r}=\frac{P c^3 r_\mathrm{em}^5\rho_\mathrm{GJ}}{B_\mathrm{NS}^2R_\mathrm{NS}^7}\Delta[(\rho/\rho_\mathrm{GJ})\times(\gamma-\gamma_0)].
  \end{equation}
Here $\gamma_0$ is the gamma-factor of particles at their birth place
(right above the polar gap) and $\gamma$ is the gamma-factor at the
emission height $r_\mathrm{em}$.  At the heights dictated by RFM, 
$(\rho/\rho_\mathrm{GJ})\times(\gamma-\gamma_0)$ must change by $\sim
10^{19}$ in order to explain $\Delta\phi=4\times10^{-3}$.  
It does not seem plausible that such drastic change in
particle energy and/or density \textit{at} the emission height
\textit{within} the mode could happen without much more significant changes in
the average profile.
 
\subsubsection{Polar gap height}
 
According to \citet{Ruderman1975}, if there is a vacuum gap between
the stellar surface and the plasma-filled magnetosphere of an aligned rotator, 
then the plasma above the gap will not co-rotate with the star itself, but will
have a slightly larger spin period.  
The discrepancy between the plasma and pulsar spin periods is proportional to the height of
the gap. 
Recently, \cite{Melrose2014} showed that non-corotating magnetospheres are also possible for the 
non-aligned rotators.
Thus, the observed lag of the pulse profile’s fiducial  point in the B-mode can
be plausibly explained by a change in the spin period of plasma above the polar cap, caused 
by slow variations in  gap height during the B-mode. As a first-order assumption, we  
estimate the necessary variation of the gap height by 
using a simple analytical equation from Appendix I in
\citet{Ruderman1975}, derived for an aligned rotator with a spherical gap. 
To explain the phase difference between the
beginning and the end of B-mode the plasma spin period must change by a fraction
$5\times10^{-7}$ and according to Eq.~A6 from \citet{Ruderman1975}:
 \begin{equation}
  \frac{P_\mathrm{end}}{P_\mathrm{beg}} = 1 + 5\times10^{-7}=\frac{R_\mathrm{NS}^2 + 3h_\mathrm{end}^2}{R_\mathrm{NS}^2 + 3h_\mathrm{beg}^2} = 1+6\frac{h\Delta h}{R_\mathrm{NS}^2},
 \end{equation}
giving that the relative change in gap height would be $\frac{\Delta h}{h}
\approx 5\%$. Since the electric potential is proportional to $h^2$, the
relative change in potential between the surface of the pulsar and the start
of the plasma-filled zone will be about 10\%.
 
Interestingly, other evidence exists that the gap potential can vary
within a B-mode instance. As was discovered before, the subpulse drift
rate slowly changes during the B-mode
\citep{Suleymanova2009,Backus2011} and is directly proportional to the
gradient of potential across the polar cap field lines
\citep{Ruderman1975,vanLeeuwen2012}.  The morphology of subpulse drift rate change
is similar to the fiducial point drift -- more rapid at the beginning
of the mode and slowing down towards B-to-Q transition
\citep{Suleymanova2009,Backus2011}.  To explain the observed change in
subpulse drift rate, the gradient of the potential must change by $-4$\%,
which is close in magnitude to the 10\% variation needed to explain the profile lag.
We must stress though that the changing subpulse drift rate and the
lag of the profile are caused by variation of the potential in
\textit{orthogonal} directions: across the field lines for the former
and along the field lines for the latter.

The variation of subpulse drift rate could also, in
principle, cause the apparent changes
in the profile shape because of the redistribution of phases of individual subpulses within the on-pulse 
window. To investigate that, we simulated a sequence of profiles from a series of drifitng
subpulses with the drift parameters taken from the single-pulse analysis of our data (Bilous et al. in prep). 
We found that the varying drift rate influenced the shape of the average profile only if the integration 
time was less than about 30 spin periods. When
integrated for at least 5 min, the average profile's shape does not
depend on the variation of the subpulse drift rate.

\section{Summary and conclusions}
\label{sec:concl}

LOFAR's observations of B0943 provided a wealth of new information about
this well-studied pulsar. The ultra-broad frequency coverage allowed us to place upper
limits on the light-travel delays and thus to constrain the emission heights in both modes. 
The long observing sessions gave the opportunity to study the gradual changes in 
the average profile during the B-mode. Finally, the high sensitivity of the telescope allowed 
us to explore the frequency evolution of the faint Q-mode profile and to compare it to 
the B-mode.

We fit the pulsar's average profile in B and Q-mode using two models. 
One of the models (``OC'') assumed that the radio emission comes from two independent 
patches in the pulsar's magnetosphere,
while the other (``MC'') served as an approximation to a cone-shaped emission region. 
For the LBA data both models worked well, resulting only in minor differences
in the values of the fitted parameters. In the HBA frequency range both B and Q-mode
profiles can be reasonably well approximated with the extension of the frequency-dependent 
LBA fit results. 

The midpoint between profile components in the LBA B-mode data appeared to
follow the $\nu^{-2}$ dispersion law down to 25\,MHz, 
allowing us to measure DM independently of the profile evolution. 
The absence of aberration and retardation  signatures in the profile 
midpoint $\phi_\mathrm{mid}(\nu)$ placed the emission region much 
closer to the stellar surface than to the light cylinder: 
$r/r_\mathrm{LC}<0.06$. If the emission in both modes comes from 
the same magnetic field lines, then at any given frequency above 20\,MHz  
radio emission heights do not differ much between the modes, with $0.94<r_\mathrm{Q}/r_\mathrm{B}<1.04$.

In the radio, B and Q-mode do not just have a different, static set of
properties, they also differ in their dynamic behaviour during a
mode. In Q-mode, the radio emission does not undergo any apparent changes
with time, whereas in the B-mode both the average profile and single
pulses evolve systematically during a mode instance \citep{Rankin2006,Suleymanova2009}.  
Thus, in addition to the rapid mode switch there exists some gradual build-up or relaxation 
in the B-mode, which, together with the unusual B-mode X-ray properties \citep[namely the potential absence of
thermal polar cap emission,][]{Hermsen2013} might be a clue to deciphering the physical trigger of the 
mode change. Our observations discover one more feature of the B-mode profile evolution: 
at 20$-$80\,MHz the midpoint between B-mode profile components is systematically shifting towards
later spin phase with the time from the start of the mode. 
The observed lag is frequency-independent down to 0.3\,ms and asymptotically changes towards a stable value of 4\,ms 
($4\times10^{-3}$ of spin phase). The lag rate is not uniform, being higher at the beginning of the mode. 
At the B-to-Q transition the profile midpoint jumps back to the earlier spin phase and remains
constant throughout Q-mode until the next Q-to-B transition.
The apparent lag of the midpoint can be interpreted as the gradual movement of 
the emission cone against the sense of pulsar rotation, with different 
field lines being illuminated over time. Another interesting explanation
is the variation of accelerating potential between the surface of the pulsar and 
the start of the plasma-filled magnetosphere above the polar gap. This explanation
connects the observed midpoint lag to the gradually evolving subpulse drift rate \citep{Rankin2006}.
The subpulse drift rate is determined by the gradient of potential in orthogonal
direction -- across the field lines in the polar cap \citep{Ruderman1975,vanLeeuwen2012}.
Both the midpoint lag and subpulse drift rate evolve similarly 
with time from the mode start: faster right after the Q-to-B transition and gradually
evening out with time. Observed magnitude of the midpoint lag points 
to about 10\% variation in the accelerating potential throughout the mode. 
At the same time the gradient of the potential
across the field lines changes by $-4$\%.

At present, there is no clear understanding of the magnetospheric
configurations during the two modes or what triggers the mode switching itself.
Most probably, our picture of pulsar emission properties in
each mode is also far from complete, being limited by 
how our line-of-sight intersects the pulsar magnetosphere.
Solving the mode-switching puzzle clearly requires much more work. One
of the obvious problems to explore is the broadband low-frequency
polarisation, both for single pulses and the average profile, as it
can provide additional important clues about the structure of the magnetic
field. Another interesting issue to be addressed is the transition
region between the modes. Any information on the time-scale of mode
switch, any special signatures found in B and Q-mode right around the
transition will give us additional insight into the mystery of the
mode switching. The ultimate goal remains to understand the
connection, if any, between the variety of emission phenomena
observed in radio pulsars.


\begin{acknowledgements}

A.V.B. thanks T.~T.~Pennucci (University of Virginia) for useful discussions. 
J.W.T.H. acknowledges funding from an NWO Vidi
fellowship and ERC Starting Grant "DRAGNET" (337062). 
LOFAR, the Low Frequency Array designed and constructed by
ASTRON, has facilities in several countries, that are owned by
various parties (each with their own funding sources), and that are
collectively operated by the International LOFAR Telescope (ILT) 
foundation under a joint scientific policy.
\end{acknowledgements}

\appendix

\section{Fitting and model applicability}
\label{subsec:fit}

\begin{figure*}[t]
   \centering
  \includegraphics[scale=0.8]{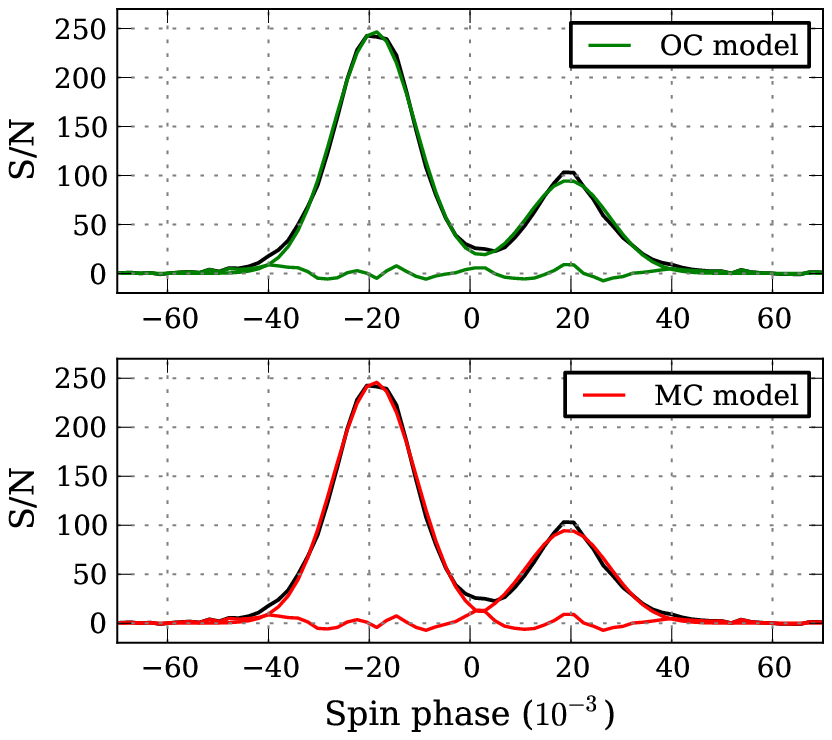}\includegraphics[scale=0.8]{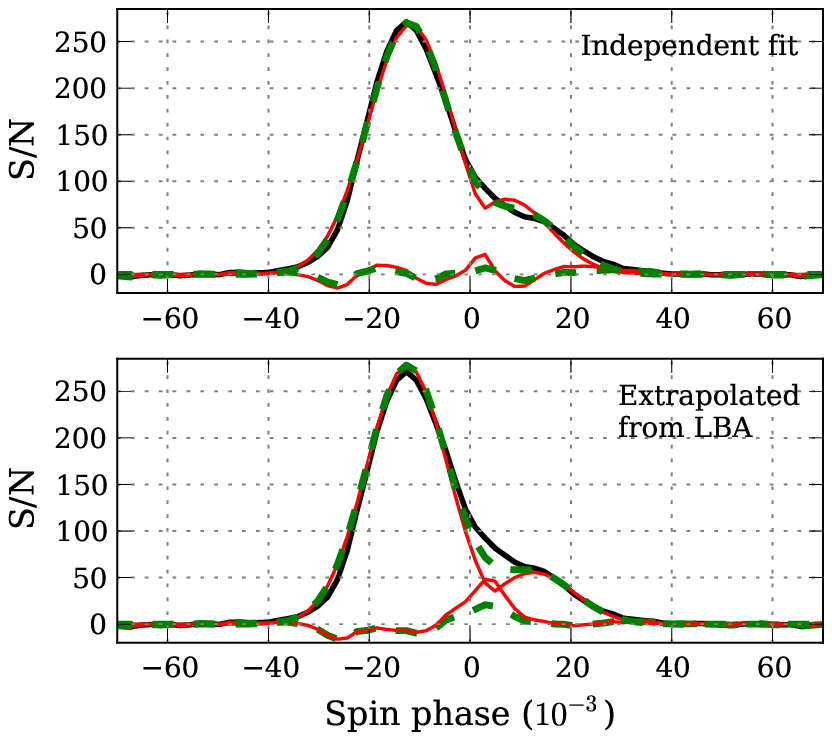}
 \caption{
\textit{Left}: an example high S/N average profile (B-mode, 2 hours of
integration, 54$-$62\,MHz), together with fit and residuals. The
residuals show a small, but statistically significant discrepancy
between the data and models. \textit{Right:} two fits of HBA B-mode
data. The upper subplot shows the LBA-like fit, with component positions,
intensities and width being a priori unconstrained.  On the lower
subplot the FWHM, ratio of component peaks and their separation are
extrapolated from the frequency dependence of corresponding LBA
values. Only the midpoint, together with the S/N of the leading component, are
fit. Neither of the fits and/or models describe the data perfectly and the
fitted values from the unconstrained fit do not agree with fitted
values extrapolated from LBA.
}
 \label{fig:modelsB}
\end{figure*}

All fits were done with the publicly available Markov-Chain Monte
Carlo software PyMC\footnote{http://pymc-devs.github.com/pymc/}.  We
assumed uniform priors on the parameters and implemented Gaussian
likelihoods.  The chains were run until they converged. The goodness
of the fit was checked by using the fitted model to simulate datasets
and then comparing the distribution of the simulated datasets to the
actual data.  For quantitative assessment we used the Bayesian
p-value\footnote{http://pymc-devs.github.io/pymc/modelchecking.html\#goodness-of-fit}
which uses Freeman-Tukey statistics to measure the discrepancy between
the data (observed or simulated) and the value predicted by the fitted
model.  The Freeman-Tukey statistics are defined only for positive
data, so we added a fixed arbitrary value to the profile, ensuring
that all data points are positive. The p-value did not depend on the
exact value of the offset added.

The Bayesian p-value is the proportion of simulated discrepancies (between
the simulated data and expected value) that are larger than their
corresponding observed discrepancies (between the observed data and
expected value).  If the model fits the data well, the p-value is
around 0.5. If the p-value is too close to 1 or 0, that indicates an
unreasonable fit. For our data, we found that the p-value depended on the
S/N of the fitted profile. If the profile had $\mathrm{S/N}\lesssim 80$ (all
Q-mode and part of B-mode), then the p-value was $0.5-0.8$. When S/N
was above 100, the p-value quickly reached 1.0, meaning that the model fit
had residuals which were larger than the noise level. If we took the
same observation of B-mode, and divided it into smaller chunks,
reducing S/N in a single chunk, the fit in a single chunk had
acceptable p-value.  An example of a high-S/N profile fit, together
with residuals, is shown in Fig.~\ref{fig:modelsB}, left. Note that
the MC model gives a slightly worse fit, underestimating the intensity
around the midpoint.

We also checked the covariances between the fitted parameters. Due to
the finite length of the MCMC chains, the measured covariance is some
random quantity around the true value (for the infinite chain), thus
even for completely independent components we expect to get some
small, but non-zero covariances between the fitting parameters. To
check when our parameters become significantly more covariant than
that, we performed the following simulation. We made two same-width
Gaussians with separation between the peaks many times exceeding the FWHM,
and fitted them in the same way as we fit our data.  We ran the
simulation 100 times, obtained a distribution for the correlation
coefficients between fit parameters and compared these distributions
to the real-data correlation coefficients.  For the B-mode, below
about 60\,MHz the correlation coefficients are compatible with those
of far-separated Gaussians. Above 60\,MHz the absolute value of
correlation coefficients starts slowly increasing and B-mode
parameters become highly covariant in the HBA data.

In general, fitting B-mode profiles in the HBA band posed some
difficulties. As in the LBA data, wherever the S/N of the average profile was
above 80 the Bayesian p-value is close to 1 and the residuals were
incompatible with noise (Fig.~\ref{fig:modelsB}, right, upper
subplot).  In addition, the fit parameters in the HBA data appeared
to be inconsistent with an extrapolation from the LBA data, both for the parameter
values and their frequency dependence within the band. The degree of
inconsistency was smaller for the OC model than for the MC model. This
discrepancy of fit results between the bands can mean that either
there is a change in parameter behaviour somewhere between our bands, 
or that our model assumptions are not correct for
higher frequencies.  To investigate the latter we constructed an
extrapolated profile, extending frequency dependence of the LBA data fit
parameters to HBA frequencies. This extrapolated profile is shown in
the bottom subplot of Fig.~\ref{fig:modelsB}, right.  Both OC and MC
models reproduce the position and width of components, but they
underestimate the emission in the region between them.  Since for the HBA data
the unconstrained fit parameters are highly correlated, this
underestimation will bias \textit{all} fit parameters.  Thus, in this
work we will not cite or discuss the fitted parameters for the HBA
B-mode data, deferring the analysis to subsequent work with a better profile
model. To first order, the average profile in HBA B-mode can be
approximated with an extrapolation of the frequency dependence of the LBA data fit
parameters.

For the Q-mode, fitted parameters were covariant at all LBA
frequencies, with correlation coefficients growing towards higher
frequencies. For the HBA data the covariance between fitted parameters gets so
large that no meaningful fit is possible without additionally
constraining the ratio of components heights. Nevertheless, simple
extrapolation of frequency dependence of LBA fit parameters works well
(see Fig.~\ref{fig:modelsB}, right).

In this work we quote for the errors on the fit parameters the 68\%
posterior probability percentiles.  When the sum, difference or ratio
of parameters is given, the error bars include the covariances between
parameters.

 \begin{figure}
   \centering
 \includegraphics[scale=0.8]{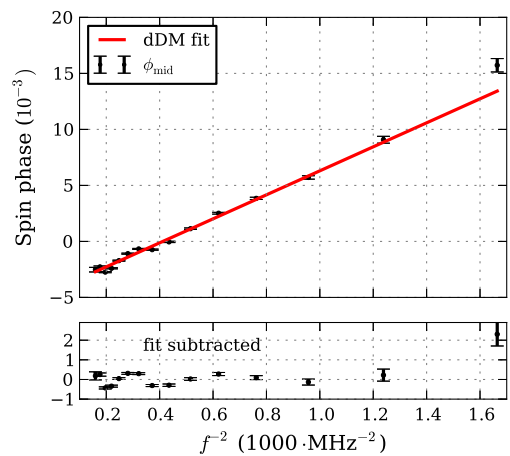}
 \caption{
The midpoint between components,
$\phi_{\mathrm{mid}}=0.5(\phi_{c1}+\phi_{c2})$ versus $\nu^{-2}$ for
one LBA B-mode observation, which was folded with some trial DM
value. The midpoint closely follows $\sim\nu^{-2}$ down to 25\,MHz and
that allows one to determine the correction to the trial DM. The extra
lag at the lowest frequency of 24.5\,MHz can be explained by
scattering delay.
}
\label{fig:DM}
\end{figure}

\section{Dispersion measure}
\label{subsec:DM}

In the LBA B-mode the midpoint between the components,
$\phi_{\mathrm{mid}} = (\phi_{c1} + \phi_{c2})/2$, is delayed with
frequency in agreement with the dispersion law:

\begin{equation}
\delta t_{\mathrm{sec}} \simeq 4.15\times10^3 \mathrm{s} \times \dfrac{\mathrm{DM_{pc/cm^3}}}{\nu^2_{\mathrm{MHz}}} 
\label{eq:DM}
\end{equation}
In other words, increasing the DM used for folding the data by $\delta
\mathrm{DM}$ shifts the midpoint between profile components by
$\delta\phi_{\mathrm{mid}}(\nu) =
(4.15\times10^3/P)\times\delta\mathrm{DM}\nu^{-2}$.  Such behaviour of
the midpoint does not hold true for all pulsars \citep{Hassall2012},
but in our case it is fortunate because it allows us to separate
interstellar 
medium (ISM) travel delay from the intrinsic profile evolution.

This dependence held true for all LBA B-mode sessions at frequencies
down to 30\,MHz.  However, at frequencies of 24$-$28\,MHz the midpoint
was lagging the $\nu^{-2}$ fit by 1$-$2\,ms (see Fig.~\ref{fig:DM} and
leftmost of Fig.~\ref{fig:stat}).  One of the plausible explanations
for this lag is scattering of the radio waves on the inhomogeneities
in the ISM.  As the result of scattering (in its simplest form), the
observer records a pulsar average profile convolved with a one-sided
exponential, $\exp(-dt/\tau_{\mathrm{scatt}})$. Scattering time
strongly depends on observing frequency ($\tau_{\mathrm{scatt}}\sim
\nu^{-4.4}$).  In our data we did not notice any prominent asymmetry
in the average profile shape, which indicates that even at the lowest
frequencies $\tau_{\mathrm{scatt}}$ is much smaller than the profile
width. However, scattering would shift the peaks of the observed
profile by an amount which is roughly (order of magnitude) comparable
to the value of $\tau_{\mathrm{scatt}}$ at that frequency. At the lowest
frequencies the components are well separated, so scattering would
shift the peaks of the both components by the same amount. The lowest
frequency in Fig.~\ref{fig:DM} is 24.5\,MHz. If we take
$\tau_{\mathrm{scatt}}(24.5 \mathrm{MHz})=1$\,ms, then scattering time
at the centre of two next subbands, 28.4 and 32.3\,MHz will be 0.5 and
0.3\,ms, within the measurement error.

The few-minute intensity variations 
of the pulsar signal during the modes (see Fig.~\ref{fig:data}) can be well explained by
 diffractive scintillation if we adopt a scattering time of 1\,ms at 24\,MHz.
Assuming the
turbulence follows a Kolmogorov spectrum and taking the values 
for the distance to the pulsar and its transverse velocity from the
ATNF catalogue v.~1.48\footnote{http://www.atnf.csiro.au/people/pulsar/psrcat/},  
the time scale for diffractive scintillation at the centre of the LBA band (50\,MHz) 
is $\Delta t_\mathrm{diff}\sim 4$\,min.

We were unable to find published measurements of scattering time for
B0943. The existing empirical recipes for scattering time give
excessively large values: 26\,ms for the empirical relationship
between scattering time and DM from \citet{Bhat2004} and 280\,ms from 
the NE2001 model of \citet{Cordes2002}.  Even
$\tau_{\mathrm{scatt}}=26$\,ms is already comparable to the width of
the average profile itself and would lead to evident distortions of
the profile shape, which contradicts our observations. However, both
of the empirical models are known to predict scattering time within
one order-of-magnitude \citep{Bhat2004,Cordes2002}. Also, NE2001 gives
the scattering time at 1\,GHz and scaling it to our low frequencies
becomes quite sensitive to the departure of turbulence from Kolmogorov
frequency dependence.  Thus we can safely conclude that our estimates
of scattering time
$\tau_{\mathrm{scatt}}(25\mathrm{MHz})\approx1$\,ms, although smaller
than predicted by empirical models, do not obviously contradict them.

For all LBA observations we folded the data with the DM measured
from the LBA B-mode midpoint, as described above, assuming that DM
does not change in Q-mode (see Table~\ref{table:data} for DM values
for each observing session).  For the HBA observation, we used the DM
from an LBA observation 2 days before.  We also looked for changes in
DM within a single observation. For all observing sessions we did not
find any deviations in DM(t) down to an error of DM(t) measurement,
$3\times10^{-4}$~pc/cm$^3$.

\bibliographystyle{aa} 
\bibliography{0943_bibliography} 

\begin{thebibliography}{36}
\expandafter\ifx\csname natexlab\endcsname\relax\def\natexlab#1{#1}\fi

\bibitem[{{Ahuja} {et~al.}(2007){Ahuja}, {Mitra}, \& {Gupta}}]{Ahuja2007}
{Ahuja}, A.~L., {Mitra}, D., \& {Gupta}, Y. 2007, \mnras, 377, 677

\bibitem[{Asgekar \& Deshpande(2001)}]{Asgekar2001}
Asgekar, A. \& Deshpande, A. 2001, \mnras, 326, 1249

\bibitem[{Backus {et~al.}(2011)Backus, Mitra, \& Rankin}]{Backus2011}
Backus, I., Mitra, D., \& Rankin, J.~M. 2011, \mnras, 418, 1736

\bibitem[{{Bhat} {et~al.}(2004){Bhat}, {Cordes}, {Camilo}, {Nice}, \&
  {Lorimer}}]{Bhat2004}
{Bhat}, N.~D.~R., {Cordes}, J.~M., {Camilo}, F., {Nice}, D.~J., \& {Lorimer},
  D.~R. 2004, \apj, 605, 759

\bibitem[{{Camilo} {et~al.}(2012){Camilo}, {Ransom}, {Chatterjee}, {Johnston},
  \& {Demorest}}]{Camilo2012}
{Camilo}, F., {Ransom}, S.~M., {Chatterjee}, S., {Johnston}, S., \& {Demorest},
  P. 2012, \apj, 746, 63

\bibitem[{{Cordes}(1978)}]{Cordes1978}
{Cordes}, J.~M. 1978, \apj, 222, 1006

\bibitem[{{Cordes} \& {Lazio}(2002)}]{Cordes2002}
{Cordes}, J.~M. \& {Lazio}, T.~J.~W. 2002, ArXiv Astrophysics e-prints

\bibitem[{{Deshpande} \& {Rankin}(2001)}]{Deshpande2000}
{Deshpande}, A.~A. \& {Rankin}, J.~M. 2001, \mnras, 322, 438

\bibitem[{{Gangadhara} \& {Gupta}(2001)}]{Gangadhara2001}
{Gangadhara}, R.~T. \& {Gupta}, Y. 2001, \apj, 555, 31

\bibitem[{{Gil} {et~al.}(1984){Gil}, {Gronkowski}, \& {Rudnicki}}]{Gil1984}
{Gil}, J., {Gronkowski}, P., \& {Rudnicki}, W. 1984, \aap, 132, 312

\bibitem[{{Hassall} {et~al.}(2012){Hassall}, {Stappers}, {Hessels}, {Kramer},
  {Alexov}, {Anderson}, {Coenen}, {et~al.}}]{Hassall2012}
{Hassall}, T.~E., {Stappers}, B.~W., {Hessels}, J.~W.~T., {et~al.} 2012, \aap,
  543, A66

\bibitem[{Hermsen {et~al.}(2013)Hermsen, Hessels, Kuiper, van Leeuwen, Mitra,
  de~Plaa, Rankin, {et~al.}}]{Hermsen2013}
Hermsen, W., Hessels, J. W.~T., Kuiper, L., {et~al.} 2013, Science, 339, 436

\bibitem[{{Karastergiou} \& {Johnston}(2007)}]{Karastergiou2007}
{Karastergiou}, A. \& {Johnston}, S. 2007, \mnras, 380, 1678

\bibitem[{{Kramer} {et~al.}(2006){Kramer}, {Lyne}, {O'Brien}, {Jordan}, \&
  {Lorimer}}]{Kramer2006}
{Kramer}, M., {Lyne}, A.~G., {O'Brien}, J.~T., {Jordan}, C.~A., \& {Lorimer},
  D.~R. 2006, Science, 312, 549

\bibitem[{{Lorimer} \& {Kramer}(2005)}]{Lorimer2005}
{Lorimer}, D.~R. \& {Kramer}, M. 2005, {Handbook of Pulsar Astronomy}, ed.
  R.~{Ellis}, J.~{Huchra}, S.~{Kahn}, G.~{Rieke}, \& P.~B. {Stetson}

\bibitem[{{Lorimer} {et~al.}(2012){Lorimer}, {Lyne}, {McLaughlin}, {Kramer},
  {Pavlov}, \& {Chang}}]{Lorimer2012}
{Lorimer}, D.~R., {Lyne}, A.~G., {McLaughlin}, M.~A., {et~al.} 2012, \apj, 758,
  141

\bibitem[{{Lyne} {et~al.}(2010){Lyne}, {Hobbs}, {Kramer}, {Stairs}, \&
  {Stappers}}]{Lyne2010}
{Lyne}, A., {Hobbs}, G., {Kramer}, M., {Stairs}, I., \& {Stappers}, B. 2010,
  Science, 329, 408

\bibitem[{{Lyne} \& {Manchester}(1988)}]{Lyne1988}
{Lyne}, A.~G. \& {Manchester}, R.~N. 1988, \mnras, 234, 477

\bibitem[{{Malofeev} {et~al.}(2000){Malofeev}, {Malov}, \&
  {Shchegoleva}}]{Malofeev2000}
{Malofeev}, V.~M., {Malov}, O.~I., \& {Shchegoleva}, N.~V. 2000, Astronomy
  Reports, 44, 436

\bibitem[{{Melrose} \& {Yuen}(2014)}]{Melrose2014}
{Melrose}, D.~B. \& {Yuen}, R. 2014, \mnras, 437, 262

\bibitem[{{Michel}(1969)}]{Michel1969}
{Michel}, F.~C. 1969, \apj, 158, 727

\bibitem[{{Mitra} \& {Deshpande}(1999)}]{Mitra1999}
{Mitra}, D. \& {Deshpande}, A.~A. 1999, \aap, 346, 906

\bibitem[{{Rankin}(1983)}]{Rankin1983}
{Rankin}, J.~M. 1983, \apj, 274, 333

\bibitem[{{Rankin}(1993)}]{Rankin1993}
{Rankin}, J.~M. 1993, \apj, 405, 285

\bibitem[{{Rankin} \& {Suleymanova}(2006)}]{Rankin2006}
{Rankin}, J.~M. \& {Suleymanova}, S.~A. 2006, \aap, 453, 679

\bibitem[{{Ruderman} \& {Sutherland}(1975)}]{Ruderman1975}
{Ruderman}, M.~A. \& {Sutherland}, P.~G. 1975, \apj, 196, 51

\bibitem[{{Shabanova} {et~al.}(2013){Shabanova}, {Pugachev}, \&
  {Lapaev}}]{Shabanova2013}
{Shabanova}, T.~V., {Pugachev}, V.~D., \& {Lapaev}, K.~A. 2013, \apj, 775, 2

\bibitem[{{Shitov}(1983)}]{Shitov1983}
{Shitov}, Y.~P. 1983, \azh, 60, 541

\bibitem[{{Stappers} {et~al.}(2011){Stappers}, {Hessels}, {Alexov}, {Anderson},
  {Coenen}, {Hassall}, {Karastergiou}, {et~al.}}]{Stappers2011}
{Stappers}, B.~W., {Hessels}, J.~W.~T., {Alexov}, A., {et~al.} 2011, \aap, 530,
  A80

\bibitem[{{Suleymanova} \& {Izvekova}(1984)}]{Suleymanova1984}
{Suleymanova}, S.~A. \& {Izvekova}, V.~A. 1984, \sovast, 28, 32

\bibitem[{{Suleymanova} {et~al.}(1998){Suleymanova}, {Izvekova}, {Rankin}, \&
  {Rathnasree}}]{Suleymanova1998}
{Suleymanova}, S.~A., {Izvekova}, V.~A., {Rankin}, J.~M., \& {Rathnasree}, N.
  1998, Journal of Astrophysics and Astronomy, 19, 1

\bibitem[{Suleymanova \& Rankin(2009)}]{Suleymanova2009}
Suleymanova, S.~A. \& Rankin, J.~M. 2009, \mnras, 396, 870

\bibitem[{{Thorsett}(1991)}]{Thorsett1991}
{Thorsett}, S.~E. 1991, \apj, 377, 263

\bibitem[{{Timokhin}(2010)}]{Timokhin2010}
{Timokhin}, A.~N. 2010, \mnras, 408, L41

\bibitem[{{van Haarlem} {et~al.}(2013){van Haarlem}, {Wise}, {Gunst}, {Heald},
  {McKean}, {Hessels}, {de Bruyn}, {et~al.}}]{vanHaarlem2013}
{van Haarlem}, M.~P., {Wise}, M.~W., {Gunst}, A.~W., {et~al.} 2013, \aap, 556,
  A2

\bibitem[{{van Leeuwen} \& {Timokhin}(2012)}]{vanLeeuwen2012}
{van Leeuwen}, J. \& {Timokhin}, A.~N. 2012, \apj, 752, 155

\end{thebibliography}

\end{document}